\documentclass[aps,prb,superscriptaddress,amsfonts,amsmath,amssymb,showpacs,floatfix,reprint,longbibliography]{revtex4-2}

\usepackage{url}
\usepackage{bm}
\usepackage{graphicx}
\usepackage{amsmath}
\usepackage{amstext}
\usepackage{amssymb}
\usepackage{amsfonts}
\usepackage{amsbsy}
\usepackage{verbatim}
\usepackage{color}
\usepackage[colorlinks=true, urlcolor=blue, linkcolor=blue, citecolor=blue, pdftex]{hyperref}
\usepackage{multirow}
\usepackage{floatrow}
\usepackage{float}
\usepackage{tikz}
\usepackage{gensymb}
\usepackage{textcomp}
\usepackage{enumitem}
\usepackage[version=3]{mhchem}
\usepackage{afterpage}
\usepackage{pbox}
\usepackage{makecell}
\usepackage{dsfont}
\usepackage{siunitx}
\usepackage{fancyhdr}
\usepackage{stackengine,wasysym,scalerel}
\usepackage{latexsym}
\usepackage{cancel}
\usepackage{array, multirow, bigdelim, makecell, booktabs}
\usepackage[misc]{ifsym}
\usepackage{fancyhdr}

\begin{document}

\title{Pinwheel valence-bond-crystal ground state of the spin-$\frac{1}{2}$ Heisenberg antiferromagnet on the \textit{\textbf{shuriken}} lattice}
\author{Nikita Astrakhantsev}
\thanks{These authors contributed equally.}
\affiliation{Department of Physics, University of Z\"urich, Winterthurerstrasse 190, 8057 Z\"urich, Switzerland}
\affiliation{Institute for Theoretical and Experimental Physics (ITEP), Moscow 117218, Russia}
\author{Francesco Ferrari}
\thanks{These authors contributed equally.}
\affiliation{Institut f\"{u}r Theoretische Physik, Goethe-Universit\"{a}t Frankfurt, Max-von-Laue-Stra{\ss}e 1, D-60438 Frankfurt am Main, Germany}
\author{Nils Niggemann}
\thanks{These authors contributed equally.}
\affiliation{Dahlem Center for Complex Quantum Systems and Institut f\"{u}r Theoretische Physik, Freie Universität Berlin, Arnimallee 14, 14195 Berlin, Germany}
\author{Tobias M\"uller} 
\affiliation{Institut f\"{u}r Theoretische Physik und Astrophysik, 
  Julius-Maximilians-Universit\"at W\"{u}rzburg, Am Hubland, D-97074 W\"{u}rzburg, Germany}
\author{Aishwarya Chauhan} 
\affiliation{Department of Physics and Quantum Centers in Diamond and Emerging Materials (QuCenDiEM) group, Indian Institute of Technology Madras, Chennai 600036, India}
\author{Augustine Kshetrimayum}
\affiliation{Dahlem Center for Complex Quantum Systems and Institut f\"{u}r Theoretische Physik, Freie Universität Berlin, Arnimallee 14, 14195 Berlin, Germany}
\affiliation{Helmholtz-Zentrum Berlin f\"{u}r Materialien und Energie, Hahn-Meitner-Platz 1, 14109 Berlin, Germany}
\author{Pratyay Ghosh}
\affiliation{Institut f\"{u}r Theoretische Physik und Astrophysik, 
  Julius-Maximilians-Universit\"at W\"{u}rzburg, Am Hubland, D-97074 W\"{u}rzburg, Germany}
\author{Nicolas Regnault}
\affiliation{Joseph Henry Laboratories and Department of Physics, Princeton University, Princeton, New Jersey 08544, USA}
\author{Ronny Thomale} 
\affiliation{Institut f\"{u}r Theoretische Physik und Astrophysik, 
  Julius-Maximilians-Universit\"at W\"{u}rzburg, Am Hubland, D-97074 W\"{u}rzburg, Germany}
\affiliation{Department of Physics and Quantum Centers in Diamond and Emerging Materials (QuCenDiEM) group, Indian Institute of Technology Madras, Chennai 600036, India}
\author{Johannes Reuther}
\affiliation{Dahlem Center for Complex Quantum Systems and Institut f\"{u}r Theoretische Physik, Freie Universität Berlin, Arnimallee 14, 14195 Berlin, Germany}
\affiliation{Helmholtz-Zentrum Berlin f\"{u}r Materialien und Energie, Hahn-Meitner-Platz 1, 14109 Berlin, Germany}
\author{Titus Neupert} 
\affiliation{Department of Physics, University of Z\"urich, Winterthurerstrasse 190, 8057 Z\"urich, Switzerland}
\author{Yasir Iqbal} 
\email[]{yiqbal@physics.iitm.ac.in}
\affiliation{Department of Physics and Quantum Centers in Diamond and Emerging Materials (QuCenDiEM) group, Indian Institute of Technology Madras, Chennai 600036, India}
  
\date{\today}

\begin{abstract}
We investigate the nature of the ground-state of the spin-$\frac{1}{2}$ Heisenberg antiferromagnet on the {\it shuriken} lattice by complementary state-of-the-art numerical techniques, such as variational Monte Carlo (VMC) with versatile Gutzwiller-projected Jastrow wave functions, unconstrained multi-variable variational Monte Carlo (mVMC), and pseudo-fermion/Majorana functional renormalization group (PF/PM-FRG) methods. We establish the presence of a quantum paramagnetic ground state and investigate its nature, by classifying symmetric and chiral quantum spin liquids, and inspecting their instabilities towards competing valence-bond-crystal (VBC) orders. Our VMC analysis reveals that a VBC with a pinwheel structure emerges as the lowest-energy variational ground state, and it is obtained as an instability of the U(1) Dirac spin liquid. Analogous conclusions are drawn from mVMC calculations employing accurate BCS pairing states supplemented by symmetry projectors, which confirm the presence of pinwheel VBC order by a thorough analysis of dimer-dimer correlation functions.
Our work highlights the nontrivial role of quantum fluctuations via the Gutzwiller projector in resolving the subtle interplay between competing orders.  
\end{abstract}

\maketitle

{\it Introduction}. The kagome lattice, which has played such a decisive role in the higher echelons of frustrated magnetism, owes much of its intriguing physics to the corner-sharing arrangement of triangular motifs. This geometry leads to only a {\it marginal} alleviation of frustration in a system of antiferromagnetically interacting spins, and in essence accounts for the appearance of novel quantum paramagnetic phases such as quantum spin liquids. In this paper, we consider the much less explored non-Archimedean~\footnote{A non-Archimedean lattice has two or more sets of topologically distinct sites. For the {\it shuriken} lattice it is two, those sites which are the vertices of squares and those which are not.} variant of the two-dimensional corner sharing arrangement of triangles, namely the {\it shuriken} lattice~\cite{Siddharthan-2001} (also referred to as the square-kagome or squagome lattice in literature) [see Fig.~\ref{fig:classical_orders}]. Following the recent experimental reporting of a gapless spin liquid in a first material realization of the {\it shuriken} geometry by spin $S=1/2$ Cu$^{2+}$ magnetic ions in KCu$_6$AlBiO$_4$(SO$_4$)$_5$Cl~\cite{Fujihala-2020}, there is renewed interest in exploring the nature of frustration induced phases in the quantum Heisenberg antiferromagnet.

\begin{figure*}[t!]
    \centering
    \begin{tikzpicture}
        \node[inner sep=0pt] at (-0.55, 0)    {\includegraphics[width=0.85\textwidth]{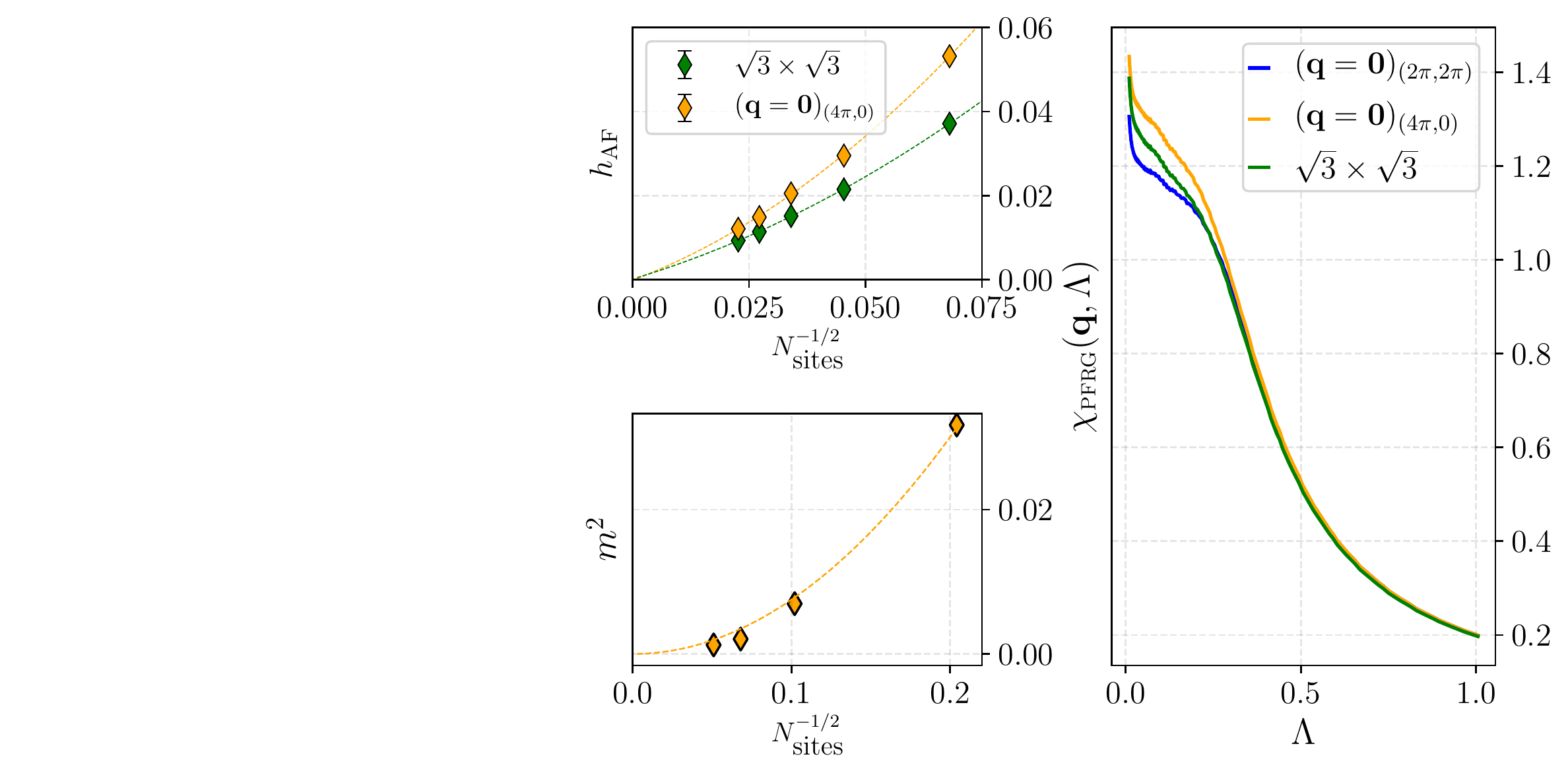}};
        \node[inner sep=0pt] at (-6.5, 0.0)    {\includegraphics[width=0.50\textwidth]{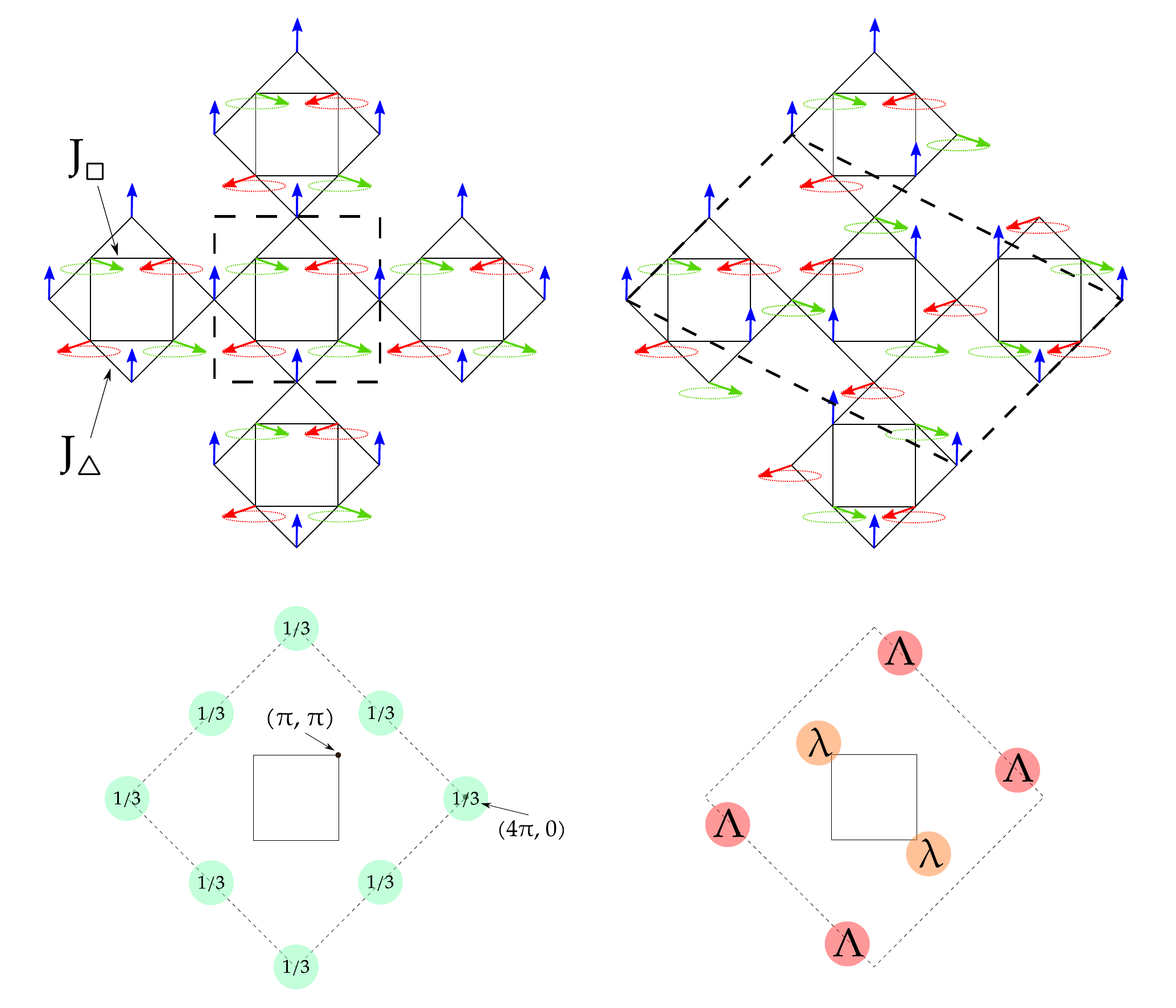}};
        \node at (-8.8, 4)  {(a) $\mathbf q = \mathbf 0$ order};
        \node at (-4.3, 4)  {(b) $\sqrt{3} \times \sqrt{3}$ order};
        \node at (4.5, 4)  {(e) PFFRG flow};
        \node at (-0.3, 4)  {(c) $h_{\rm AF}$ parameter (VMC)};
        \node at (-0.1, -0.05)  {(d) Magnetization (mVMC)};
    \end{tikzpicture}
    \caption{(a)-(b) Top row: Illustration of two types of magnetic orders within the ground state manifold of the classical Heisenberg antiferromagnet on the {\it shuriken} lattice~\cite{Richter-2009,Rousochatzakis-2013}. The two symmetry inequivalent nearest-neighbor bonds are labelled as $J_{\bigtriangleup}$ and $J_{\square}$. The $\mathbf{q}=\mathbf{0}$ ($\sqrt{3}\times\sqrt{3}$) order has an angle of $120\degree$ between neighboring spins, and a magnetic unit cell which is identical (three-times enlarged) compared to the six-site geometrical unit cell is marked by a dashed line in (a). The green ellipses depict further degrees of freedom present in the classically degenerate ground state manifold. Bottom row: The first (solid line) and extended (dashed line) Brillouin zones of the {\it shuriken} lattice showing the location of the Bragg peaks with the fraction of the total spectral weight of the classical (a) $\mathbf{q}=\mathbf{0}$ order, at $(4\pi,0)$ and $(2\pi,2\pi)$ [and symmetry related points] with equal spectral weight, and (b) $\sqrt{3}\times\sqrt{3}$ order, at $(2\pi \pm q, 2\pi \mp q)$ with $q = 4\pi/3$, and leading subdominant peaks at $(q,-q)$ with 36\%, i.\,e., $\lambda/\Lambda=0.36$, of the spectral weight of the dominant ones~\cite{note_peaks}. The $\sqrt{3}\times\sqrt{3}$ order breaks the four-fold rotational symmetry. From (c) VMC, the size scaling of the $h_{\rm AF}$ parameter (fictitious Zeeman field~\cite{supp_mat}) for $\mathbf{q}=\mathbf{0}$ and $\sqrt{3}\times\sqrt{3}$ orders on finite clusters of $N_{\rm sites}=6L^2$ sites, with $L=6,9,12,15$ and $18$ (quadratic fit), (d) mVMC, the size-scaling of the sublattice magnetization for the $\mathbf{q}=\mathbf{0}$ order finite clusters of $N_{\rm sites}=6L^2$ sites, with $L=2,4,6$ and $8$ (quadratic fit), and (e) PFFRG, the RG flow of the susceptibility tracked at the dominant ordering vectors of the two classical orders.}
    \label{fig:classical_orders}
\end{figure*}
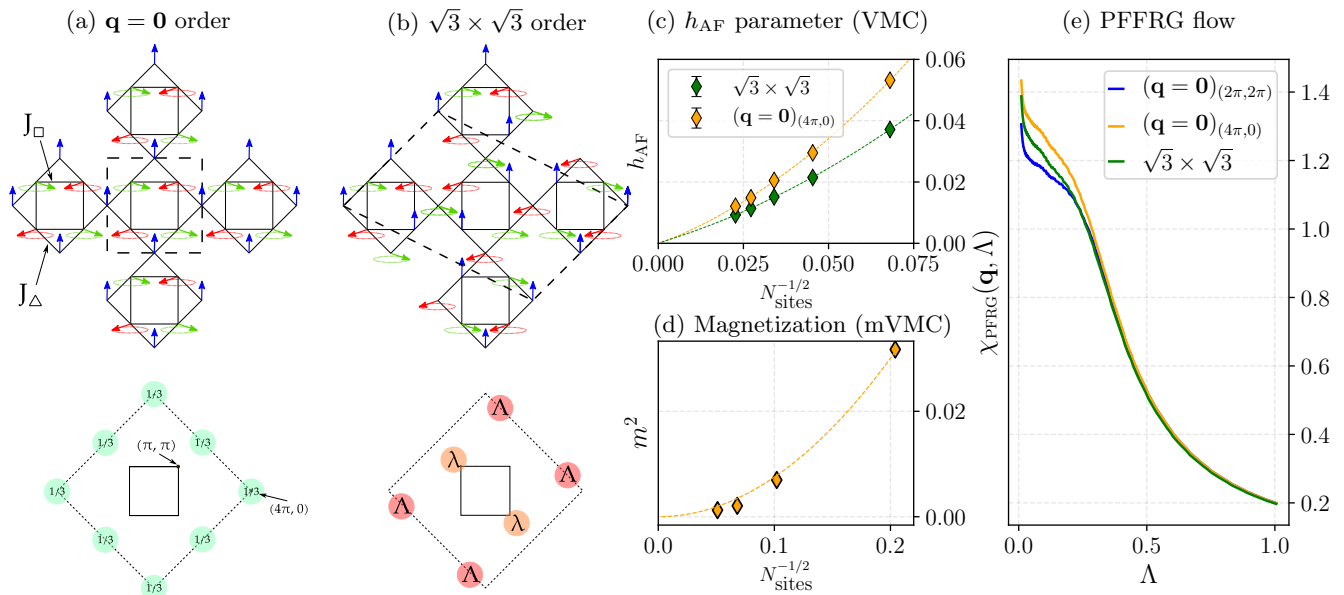

In this work, we investigate the ground state of the Heisenberg antiferromagnetic model
\begin{equation}
    \hat{{\cal H}} = J \sum_{\langle i,j \rangle} \mathbf{\hat S}_{i} \cdot \mathbf{\hat S}_{j} \label{eqn:heisenberg}
\end{equation}
for $S=1/2$ operators $\hat{\mathbf{S}}_{i}=(\hat{S}^x_{i},\hat{S}^y_{i},\hat{S}^z_{i})$ decorated on the {\it shuriken} lattice, where the two symmetry inequivalent exchange couplings, i.e., on the square ($J_{\square}$) and triangular ($J_{\bigtriangleup}$) bonds (see Fig.~\ref{fig:classical_orders}(a)) are taken as equal and denoted by $J$ (=$J_{\bigtriangleup}=J_{\square})$~\cite{supp_mat}. We employ (i) Variational Monte Carlo (VMC) with Gutzwiller-projected fermionic wave functions on large system sizes (up to $2400$ sites), (ii) many-variable Variational Monte Carlo (mVMC) involving an unconstrained optimization of a BCS pairing function on system sizes up to $384$ sites, and (iii) pseudo-fermion functional renormalization group (PFFRG) analysis to firmly establish the absence of long-range magnetic order in the ground state of~\eqref{eqn:heisenberg}, which remained to be conclusively established within exact diagonalization~\cite{Richter-2009,Nakano-2013,Rousochatzakis-2013,Morita-2018,Hasegawa-2018}, mean-field~\cite{Lugan-2019}, and perturbative~\cite{Tomczak-2003,Ralko-2015} schemes. To identify the precise nature of the quantum paramagnetic ground state, we construct symmetric and chiral U(1) fermionic mean-field Ans\"atze of quantum spin liquids, compute their projected energies, and investigate their potential instability towards competing VBC orders which have been proposed within a quantum dimer model approach~\cite{Rousochatzakis-2013,Ralko-2015} and a large-$N$ analysis~\cite{Siddharthan-2001}. Our study finds an instability of the U(1) Dirac spin liquid towards a pinwheel (PW) VBC order with a $2\times2$ expanded, i.\,e., $24$-site unit cell ~\cite{Rousochatzakis-2013,Ralko-2015} which emerges as the lowest energy variational state, in contrast to the findings in Ref.~\cite{Ralko-2015} which claimed for the stabilization of a Loop-6 (L6) VBC, see also Fig.~\ref{fig:energies_pwvsl6}. An unconstrained mVMC optimization of the BCS pairing function starting from a random choice of pairing amplitudes is also found to converge to a long-range dimer ordered ground state with a PW-VBC type structure, as revealed from the dimer-dimer correlation functions. These findings are further corroborated by a PFFRG analysis of the dimer response functions. The estimates of the ground state energy on finite-clusters obtained within VMC by the application of a couple of Lanczos steps to the PW-VBC state supplemented by a zero-variance extrapolation~\cite{sorella2001,becca2015}, as well as those obtained from mVMC are found to be in excellent agreement. 
The respective estimates in the thermodynamic limit obtained by finite-size scaling are in good agreement with those obtained from our infinite projected entangled pair state (iPEPS) and pseudo-Majorana functional renormalization group (PMFRG) calculations, thus lending strong evidence in favor of a PW-VBC ground state of the $S=1/2$ Heisenberg antiferromagnet on the {\it shuriken} lattice. 

%%%%%%%%%%%%%%%%%%%%%%%%%%%%%%%%%%%%%%%%%%%%
\floatsetup[table]{capposition=top}
\begin{table}
\centering
\begin{tabular}{llccc}
 \hline \hline
    \multicolumn{1}{c}{Ansatz}
    & \multicolumn{1}{l}{Fluxes}
    & \multicolumn{1}{c}{MF energy}
    & \multicolumn{1}{c}{Proj. energy}
    & \multicolumn{1}{c}{MF spectrum}
 \\ 
\hline \hline 
\multirow{4}{*}{FS} &
 $(0,0,0)$ & $-0.36570$ & $-0.42714(2)$ & Fermi surface \\
 & $(0,0,\pi)$ & $-0.38388$ & $-0.41720(3)$ & Dirac points\\
 & $(\pi,0,0)$ & $-0.36657$ & $-$ & Flat band \\
 & $(\pi,0,\pi)$ & $-0.37130$ & $-0.41362(3)$ & Fermi surface \\ \\
 
 \multirow{4}{*}{Chiral} &
 ($\frac{\pi}{2},\frac{\pi}{2},0)$ & $-0.35041$ & $-$ &  Flat band \\ 
& ($\frac{\pi}{2},\frac{\pi}{2},\pi$) & $-0.39803$ & $-0.40489(3)$ &  Gapped \\ 
& $(\frac{\pi}{2},-\frac{\pi}{2},0)$ & $-0.38040$ & $-0.38702(4)$ & Fermi surface \\ 
& ($\frac{\pi}{2},-\frac{\pi}{2}$,$\pi$) & $-0.36123$ & $-0.40205(3)$ & Fermi surface \\ \\
 
 \multirow{2}{*}{Dimer} 
 & L6-VBC & $-0.41013$ & $-0.43009(1)$ & Gapped \\
 &  PW-VBC & $-0.40623$ & $-0.43333(1)$ & Gapped \\
\hline \hline
\end{tabular}
\caption{For Hamiltonian~\eqref{eqn:heisenberg}, we present the mean-field (MF) and Gutzwiller projected (proj) ground-state energy per site (in units of $J$) on the $4\times 4\times 6$ cluster for the different fully symmetric (FS) and chiral U(1) quantum spin liquid Ans\"atze [see Fig.\,S4 of~\cite{supp_mat}] labelled by the flux triad $(\Phi_{\bigtriangleup},\Phi_{\displaystyle\triangleright},\Phi_{\square})$ [see text], as well as dimer states. The $(\pi,0,0)$ and $(\pi,0,\pi)$ Ansatz have extensively degenerate levels at half-filling which prevents a computation of their energy (as the wave function cannot be uniquely defined), and are thus marked by ``$-$''.}
\label{tab:en-j=1}
\end{table}
%%%%%%%%%%%%%%%%%%%%%%%%%%%%%%%%%%%%%%%%%%%

{\it Results}. We start by employing fermionic VMC to investigate the possible presence of magnetically ordered ground states with two different periodicity, (i) a translationally invariant, i.\,e., $\mathbf{q}=\mathbf{0}$ state [see Fig.~\ref{fig:classical_orders}\,(a)] and (ii) the so-called $\sqrt{3}\times\sqrt{3}$ state [see Fig.~\ref{fig:classical_orders}\,(b)]. Details on the form of the variational wave functions are given in the Supplemental Material~\cite{supp_mat}. The Zeeman field variational parameter, $h_{\rm AF}$, extrapolates to zero in the thermodynamic limit [see Fig.~\ref{fig:classical_orders}\,(c)], indicating the absence of long-range magnetic order in the ground state. Additional evidence is provided by mVMC calculations, in which the sublattice magnetization $m^{2}$ is computed by evaluating the spin-spin correlation $\langle\mathbf{\hat S}_{i}\cdot \mathbf{\hat S}_{j}\rangle$ at maximum distance (for two spins $i,j$ within the same sublattice)~\cite{sandvik1997},
where $\langle \cdots \rangle$ denotes the expectation value over the variational state $|\phi_{\rm pair}\rangle$~\cite{supp_mat}. The sublattice magnetization is seen to display a similar size scaling as the $h_{\rm AF}$ parameter of VMC [see Fig.~\ref{fig:classical_orders}\,(d)], thus confirming the absence of magnetic order. These results are further corroborated by a PFFRG analysis~\cite{supp_mat} which does not find in the RG flow any evidence for a divergence or a breakdown of the susceptibility at the ordering wave vectors of either the $\mathbf{q}=\mathbf{0}$ or $\sqrt{3}\times\sqrt{3}$ orders [see Fig.~\ref{fig:classical_orders}\,(e)].

Thus, having established the paramagnetic character of the ground state, we proceed towards deciphering its nature. To this end, we construct a family of fully symmetric and chiral fermionic mean-field Ans\"atze with a U(1) gauge structure based on a symmetry classification of flux patterns. These Ans\"atze are completely determined by specifying the fluxes threading three distinct plaquettes within the unit cell of the lattice: vertically oriented triangles $(\Phi_{\bigtriangleup})$, horizontally 
oriented triangles $(\Phi_{\displaystyle\triangleright})$, and squares $(\Phi_{\square})$. Henceforth, we label the mean-field Ans\"atze by 
specifying the flux triad $(\Phi_{\bigtriangleup},\Phi_{\displaystyle\triangleright},\Phi_{\square})$. 
We obtain four distinct fully symmetric U(1) spin liquids (see Table~\ref{tab:en-j=1} and \cite{supp_mat}), and several chiral states out of which we focus only on those with a $\pi/2$ flux through triangles. In Table~\ref{tab:en-j=1}, we present the energies of the (self-consistent) mean-field and Gutzwiller-projected wave functions for these states. At the mean-field level, the chiral $(\frac{\pi}{2},\frac{\pi}{2},\pi)$ state has the lowest energy, in complete compliance with the Rokhsar rules~\cite{Rokhsar-1990}~\footnote{The Rokhsar rules dictate that the optimal state at the mean-field level is given by a flux pattern where polygons with an odd number of sides (here, triangles) prefer to have a $\pi/2$ flux in order to minimize the fermion energy, while polygons with $4n$ sides (here, squares), where $n$ is an integer, prefer to have $\pi$ flux, and finally polygons with $4n+2$ sides (the hexagons) prefer zero flux.}. However, after Gutzwiller projection, the chiral spin liquid is no longer energetically competitive, and the $(0,0,0)$ flux uniform Ansatz featuring a spinon Fermi surface (SFS) emerges as the lowest energy spin-liquid state, followed by the $(0,0,\pi)$ state which is a U(1) Dirac spin liquid (DSL). Since, in two spatial dimensions, the U(1) SFS and DSL are potentially susceptible to gap opening instabilities~\cite{Hermele-2004,Song-2019}, we investigate their instability towards previously proposed VBC candidates~\cite{Rousochatzakis-2013,Ralko-2015}.

\begin{figure}[t!]
\vspace{-0.6cm}
    \centering
    \begin{tikzpicture}
        \node[inner sep=0pt] at (0, 0.6)    {\includegraphics[width=1.0\textwidth]{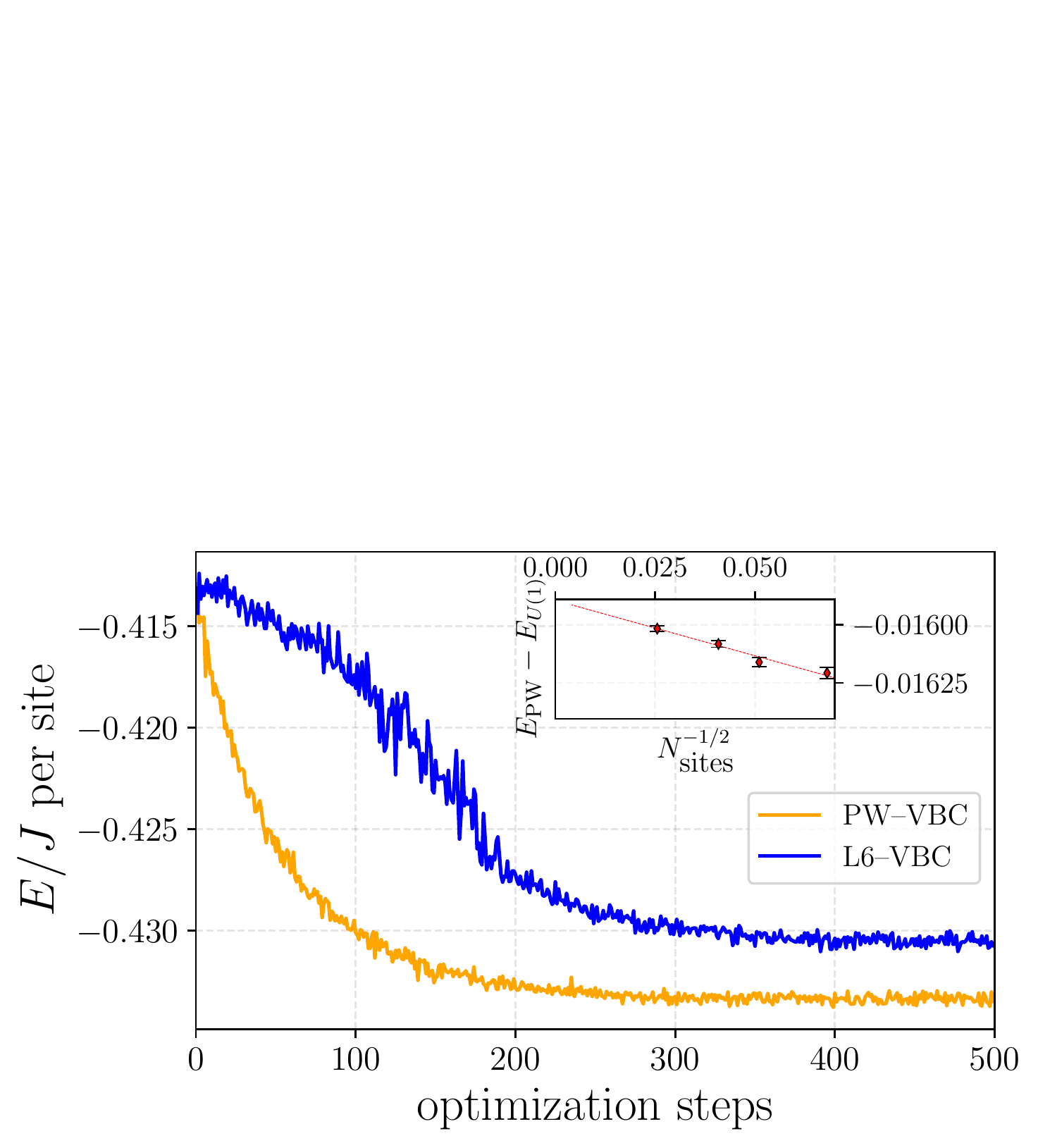}};
        \node[inner sep=0pt] at (0.5, 2.9)    {\includegraphics[width=0.75\textwidth]{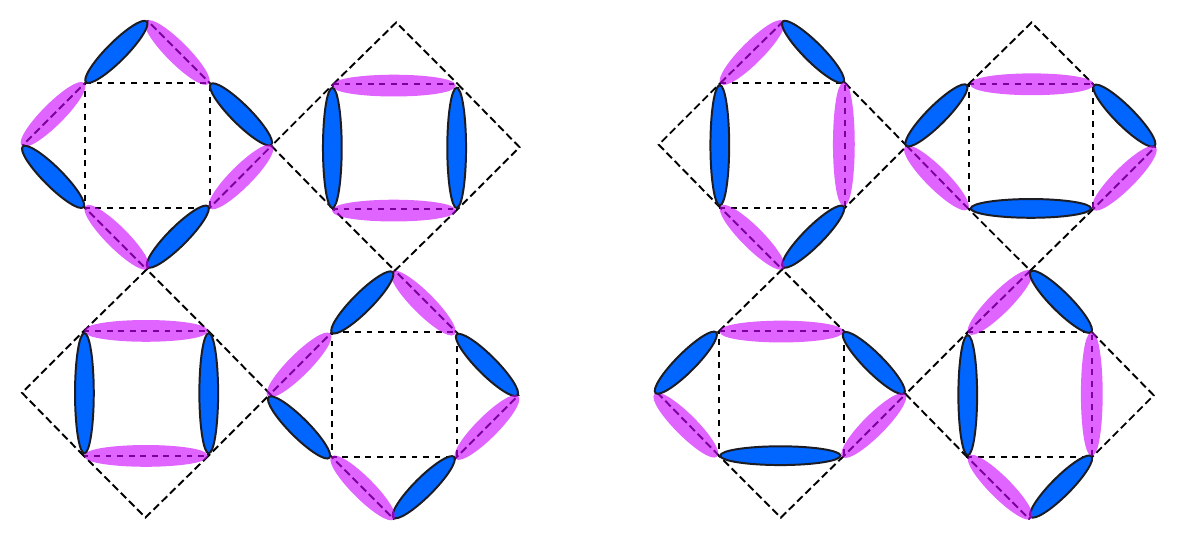}};
        \node at (-1.3, 4.6)  {(a) Pinwheel VBC};
        \node at (2.2, 4.6)  {(b) Loop-6 VBC};
        \node at (-1.0, 1.1)  {(c) Energies optimization};
    \end{tikzpicture}
    \caption{Two competing dimer orders, (a) PW-VBC and (b) L6-VBC on the \textit{shuriken} lattice. (c) The evolution of the energy per site during a typical VMC optimization for the VBCs, here shown for Hamiltonian~\eqref{eqn:heisenberg} on the $L=20$ cluster. The inset shows the finite-size scaling of the energy gain of the PW-VBC state with respect to the U(1) DSL.}
    \label{fig:energies_pwvsl6}
\end{figure}
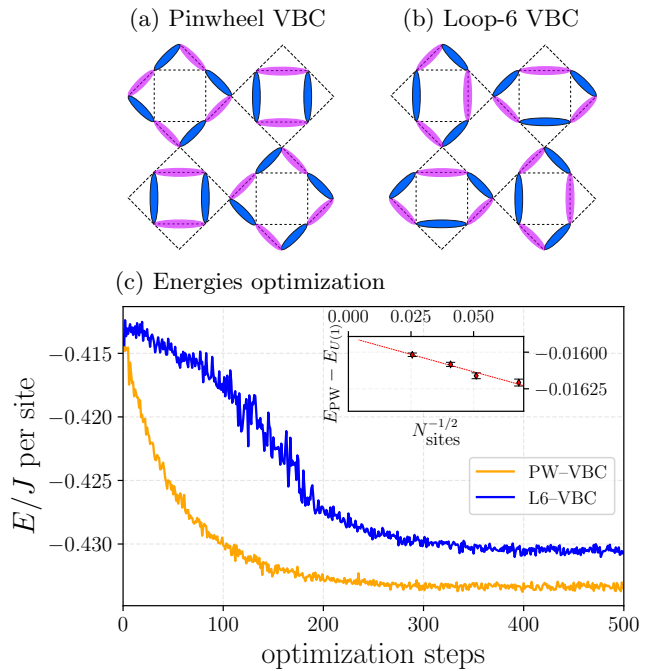

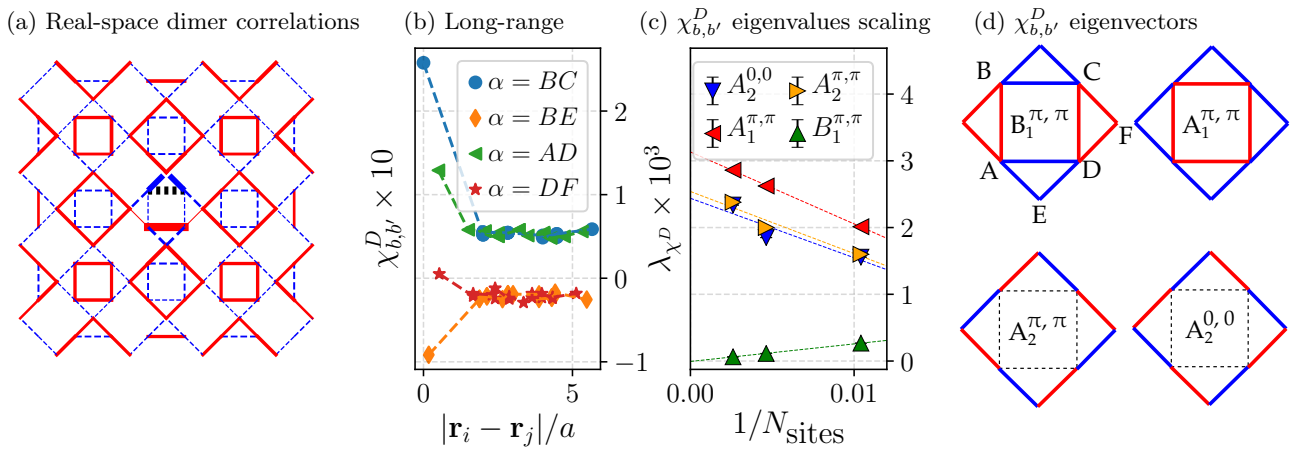
\begin{figure*}[t!]
    \centering
    \begin{tikzpicture}
        \node[inner sep=0pt] at (0, 0)    {\includegraphics[width=1.\textwidth]{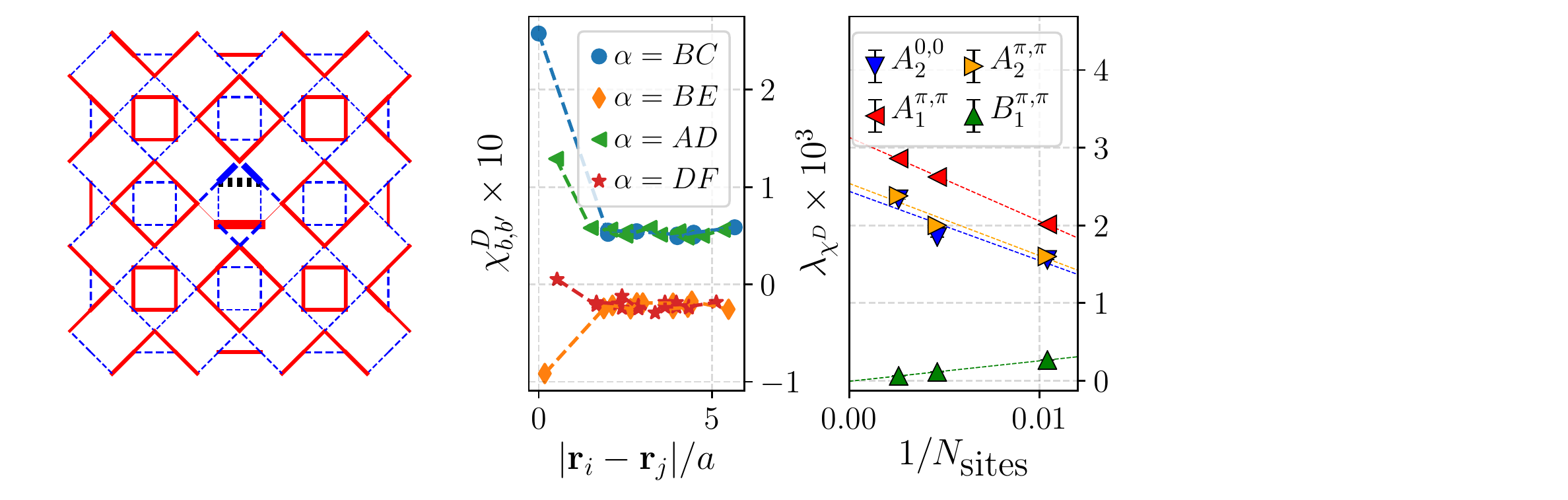}};
        
        \node[inner sep=0pt] at (6.5, 0.3)    {\includegraphics[width=0.3\textwidth]{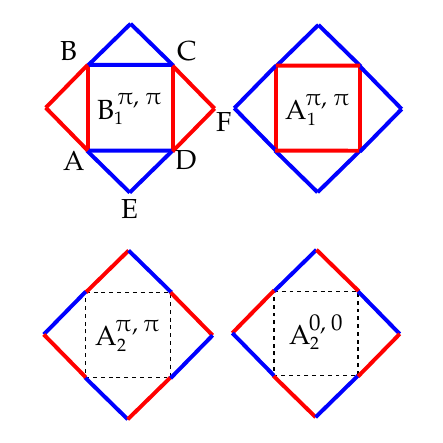}};
        \node at (-6.0, 3)  { (a) Real-space dimer correlations};
        \node at (-2, 3)  { (b) Long-range};
        \node at (2, 3)  { (c) $\chi^D_{b,b'}$ eigenvalues scaling};
        \node at (6, 3)  { (d) $\chi^D_{b,b'}$ eigenvectors};
    \end{tikzpicture}
    \caption{(a)~Dimer-dimer correlations between the base bond (between sublattice B and C sites~\cite{supp_mat}), marked by a black dotted line, and other bonds measured within mVMC on the $L=8$ cluster. 
    %(for typographical reasons $\chi_{b,b'}^{D}$ is multiplied by $10$ and rounded to 2 digits). 
    Solid red (dashed blue) lines represent positive (negative) correlation values. An analogous figure with the base bond on the side of a triangle can be found in~\cite{supp_mat}. (b) Long-range correlations between the base bond and other bonds labeled by $\alpha$. Only correlations between the $(0,\,0)$ and $(2 n,\,2 m)$ unit cells are shown due to the alternating $2 \times 2$ pattern in bond correlations. (c) Infinite-volume extrapolation of the largest eigenvalues of $\chi_D$. (d) The dominant eigenvectors of $\chi^D$: red bonds are positive, while blue are negative and other bonds are zero. In the middle of each drawing, the high-symmetry momentum of the representation is shown, as well as the corresponding irreducible representation label of the $D_4$ point symmetry group.}
    \label{fig:dimer_main}
\end{figure*}

A quantum dimer model treatment of \eqref{eqn:heisenberg}, truncated to a minimal nearest-neighbor valence bond basis, identified a PW-VBC with loop-4 resonances~\cite{Rousochatzakis-2013} [see Fig.~\ref{fig:energies_pwvsl6}\,(a)]. This picture was subsequently challenged in Ref.~\cite{Ralko-2015} by a L6-VBC with loop-6 resonances [see Fig.~\ref{fig:energies_pwvsl6}\,(b)] when accounting for a basis beyond nearest-neighbors. In particular, it was argued that the virtually excited long-range singlets that are induced around defect triangles lead to an enhancement of loop-6 resonances compared to loop-4, helping stabilize the L6-VBC \textit{in lieu} of the PW-VBC~\cite{Ralko-2015}. Here, we investigate the energetic competition between PW-VBC and L6-VBC orders within VMC and mVMC wherein the effect of quantum fluctuations is captured via the Gutzwiller projector. To construct variational VBC states within VMC, we consider each of the symmetric spin-liquid Ans\"atze listed in Table~\ref{tab:en-j=1}, and allow the hopping amplitudes to take different values according to the dimer pattern of the strong/weak symmetry inequivalent bonds within the $24$-site VBC unit cells~\footnote{The PW-VBC (L6-VBC) structure has six (eight) distinct symmetry inequivalent bonds/hopping amplitudes, and upon fixing one of them as the reference, we are left with five (seven) amplitudes which we optimize~(see also \cite{supp_mat})}.
Our study reveals that while the SFS spin liquid remains robust to both these VBC perturbations, the U(1) Dirac spin liquid destabilizes towards both these VBCs, with the PW-VBC yielding a lower energy [see Fig.~\ref{fig:energies_pwvsl6}\,(c) and Table~\ref{tab:en-j=1}]. It is interesting to note that at the self-consistent mean-field level the L6-VBC has a lower energy compared to the PW-VBC, but the relative hierarchy is inverted in favor of the PW-VBC once the Gutzwiller projector is enforced within VMC, highlighting the role of quantum fluctuations in resolving the delicate competition of these two dimerized states. Furthermore, VMC calculations show that the energy gain of the PW-VBC relative to the U(1) Dirac spin liquid remains finite in the thermodynamic limit [see inset of Fig.~\ref{fig:energies_pwvsl6}\,(c)] indicating the size-consistency of the PW-VBC state, thereby lending support to it being a stable variational ground state in the thermodynamic limit. The energy of the PW-VBC Ansatz is found to be lower compared to the SFS spin-liquid state, thus representing the optimal wave function within our VMC calculations.

To obtain competitive wave functions within mVMC, we impose a $2 \times 2$ unit cell periodicity on the parameters of the variational state $|\phi_{\rm pair}\rangle$~\cite{Misawa_2016,supp_mat}. The properties of the optimized wave function are assessed by measuring the dimer-dimer correlation function $\chi^D_{b,b'} = \langle \hat{D}_b \hat{D}_{b'} \rangle - \langle \hat{D}_b \rangle \langle \hat{D}_{b'} \rangle$ for all pairs of bonds in the system, $0 \leqslant b,\,b' < N_{\mbox{\small bonds}}$, where $\hat{D}_b=\mathbf{\hat S}_{i} \cdot \mathbf{\hat S}_{j}$, with $i,\,j$ being sites at ends of the bond $b$. In Fig.~\ref{fig:dimer_main}\,(a), we show the dimer-dimer correlations between the base (square) bond and other bonds lying within few unit cells, which display the characteristic pinwheel structure found also within VMC. To carry on a quantitative assessment of the VBC character of the ground state, we need to define suitable scalar order parameters to perform an infinite-volume extrapolation of the dimer order. Thus, we regard $\chi^D_{b,b'}$ as a matrix in the bond indices and we diagonalize it; the resulting set of eigenvalues/eigenvectors pairs (${\lambda, \ A^\lambda_b}$) is used to define the operators $\hat{\mathcal{O}}_\lambda = \sum_b A^\lambda_b \hat D_b$, each of them corresponding to a certain momentum and irreducible representation of the lattice point group. The tendency to establish a finite expectation value of one of these operators, and thus spontaneously break the corresponding lattice symmetry, is measured by the susceptibility $\chi_{\hat{\mathcal{O}}_\lambda} = \langle \hat{\mathcal{O}}_\lambda^{\dagger} \hat{\mathcal{O}}_\lambda \rangle -
\langle \hat{\mathcal{O}}_\lambda^{\dagger} \rangle\langle \hat{\mathcal{O}}_\lambda \rangle=\lambda$ extrapolated to thermodynamic limit~\cite{astrakhantsev2021brokensymmetry}.

This extrapolation requires knowledge of the order parameter scaling law. We argue that $1 / L^2$, i.\,e., inverse-volume scaling is a suitable choice. To check that, in Fig.~\ref{fig:dimer_main}\,(b) we show long-range behavior of correlations between the base bond ($BC$ in the $(0,\,0)$ unit cell) and bonds in the other unit cells with bond within unit cell labeled by $\alpha$ (see drawing in Fig.~\ref{fig:dimer_main}\,(c)). We observe that (i) the correlator saturates almost immediately with distance, suggesting exponential decay finite-range corrections to correlations, and (ii) correlations between the base bond and bonds with different $\alpha$ converge to {\it different and finite} values, which paves the way to finite susceptibility $\chi_{\hat{\mathcal{O}}_\lambda}$ for some symmetry-breaking operator $\hat{\mathcal{O}}_\lambda$.

In the illustrations shown to the left and right of Fig.~\ref{fig:dimer_main}\,(c) we depict the leading eigenvectors of $\chi^D_{b,b'}$ and show the infinite-volume extrapolation of the corresponding eigenvalues $\lambda$  (i.\,e., of the susceptibilities $\chi_{\hat{\mathcal{O}}_\lambda}$). We observe that the susceptibilities corresponding to the $A_{1}^{\pi,\pi}$ and $A_{2}^{\pi,\pi}$ irreducible representations, which are reflective of the PW-VBC phase symmetry structure, clearly extrapolate to finite values, while, the $B_{1}^{\pi,\pi}$ irreducible representation susceptibility indicating L6-VBC phase structure is found to vanish (within error-bars) in the thermodynamic limit. 

We also probe, within PFFRG, the tendency towards PW-VBC and L6-VBC symmetry breaking patterns, and observe that although the dimer-response functions for these two orders get enhanced under RG flow indicating dimerization, they are of similar magnitude~\cite{supp_mat}. The absence of a categorical identification of the dimerization tendency within PF-FRG is rooted in the fact that one does not take into account higher-point vertex functions which ultimately seem to prove decisive in accurately resolving the delicate competition between the two VBC candidates. 

\begin{figure}
    \includegraphics[width=0.95\textwidth]{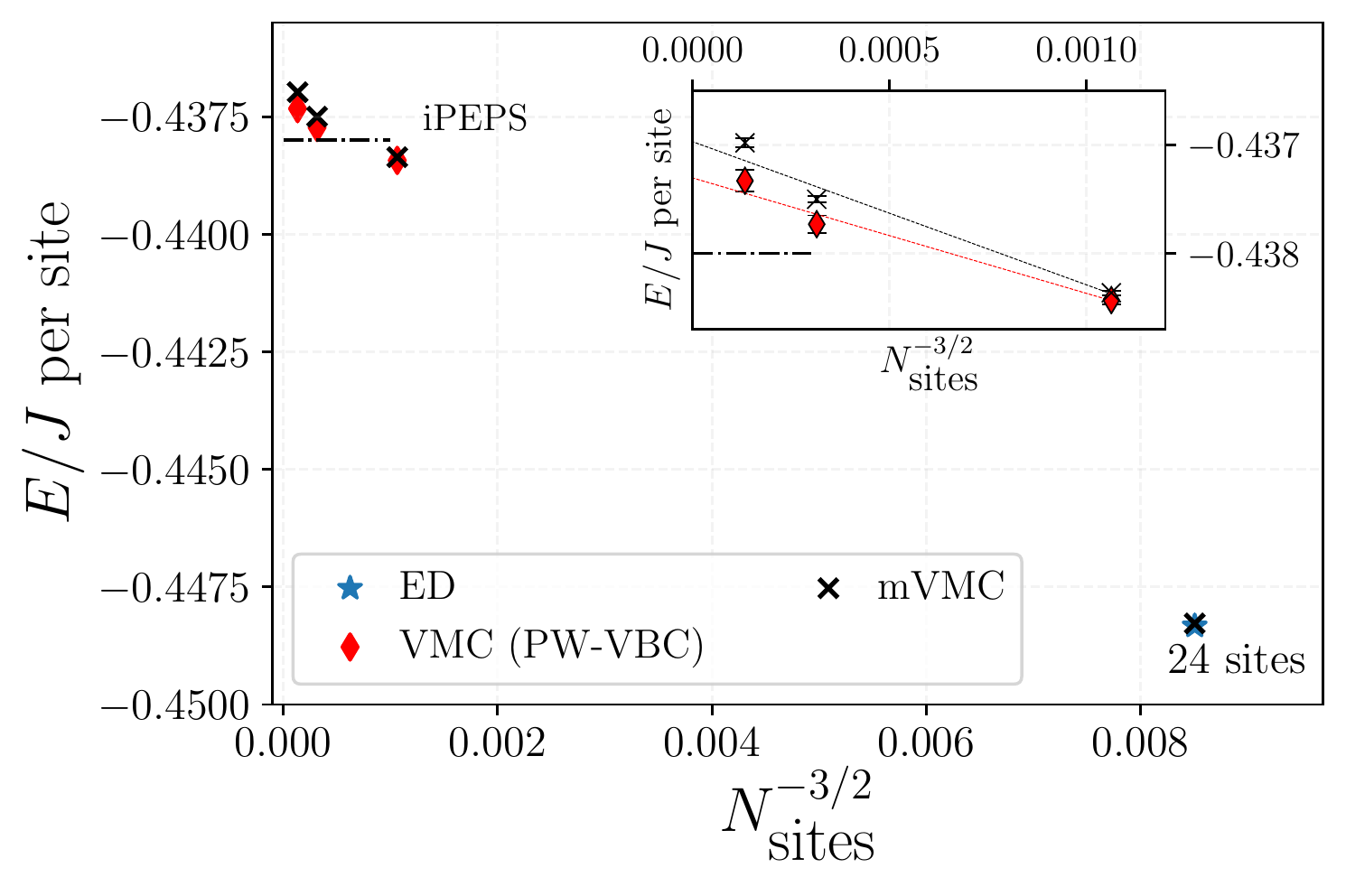}
    \caption{The ground state energies on the $L=4,\,6$ and $8$ clusters obtained from VMC and mVMC together with the inset showing the finite-size scaling. The blue star denotes the ED energy on the $D_{4}$ symmetric $6 \times 2^2$ cluster, the iPEPS energy (obtained by a quadratic fit for the three largest bond dimensions) is marked by horizontal lines~\cite{supp_mat}.}
    \label{fig:energies_all}
\end{figure}

Having established, from both VMC and mVMC, that the ground state of the system possesses long-range dimer order, we discuss other static quantities, namely the extrapolation of the ground state energy, reported in Fig.~\ref{fig:energies_all}, and the equal-time spin structure factor, shown in Fig.~\ref{fig:Sq_8x8}. Within VMC, an improved estimate of the ground state energy on finite clusters can be achieved by applying a few Lanczos steps to the variational state (here the PW-VBC), and performing a zero-variance extrapolation~\cite{sorella2001,becca2015,Iqbal-2013,Iqbal-2014_gap,Iqbal-2016_tri,Iqbal-2018_breathing}. The resulting estimate of the ground state energy is found to be equal (within three error-bars) with the mVMC energies on the $L=4$, $6$, and $8$ clusters. Furthermore, the finite-size-scaling estimate of the thermodynamic ground state energy from VMC and mVMC
\begin{equation}
    E^{\infty}_{\rm mVMC}=-0.43696(17), \quad E^{\infty}_{\rm VMC}=-0.43730(13)
\end{equation}
are equal within two error-bars and in excellent agreement with that obtained from iPEPS and consistent with PMFRG at finite temperature directly in the thermodynamic limit. Finally, the (equal-time) static spin structure factor $S({\bf q})$~\cite{supp_mat} for the PW-VBC ground state obtained from both VMC and mVMC approaches are shown in Fig.~\ref{fig:Sq_8x8}. One observes a diffused distribution of intensity along the extended Brillouin zone boundaries. We observe that within VMC the estimate of the ground state $S({\bf q})$ obtained by applying two Lanczos steps on the bare PW-VBC wave function displays soft maxima in close vicinity to the pinch points [Fig.~\ref{fig:Sq_8x8}(a)] seen in a large-$N$ analysis [Fig. S2 of~\cite{supp_mat}] in conformity with mVMC [Fig.~\ref{fig:Sq_8x8}(b)].

\begin{figure}[t]
    \centering
    \begin{tikzpicture}
        \node[inner sep=0pt] at (-0.15, 0)    {\includegraphics[width=0.475\textwidth]{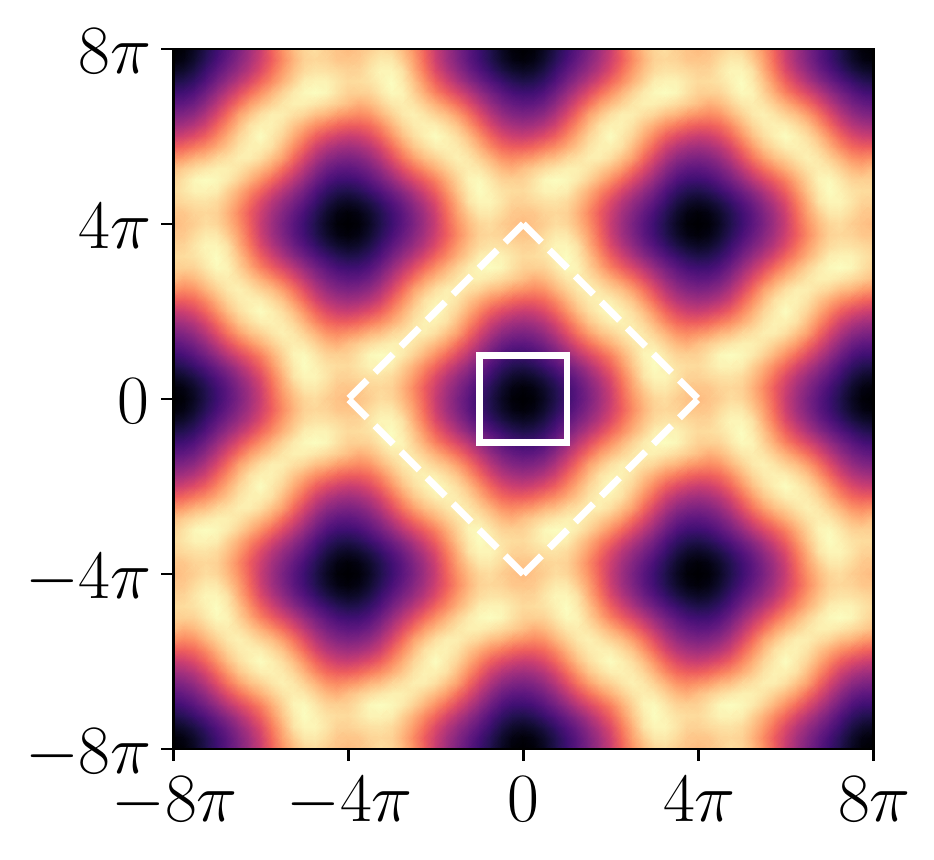}};
        \node[inner sep=0pt] at (4.15, 0)    {\includegraphics[width=0.53\textwidth]{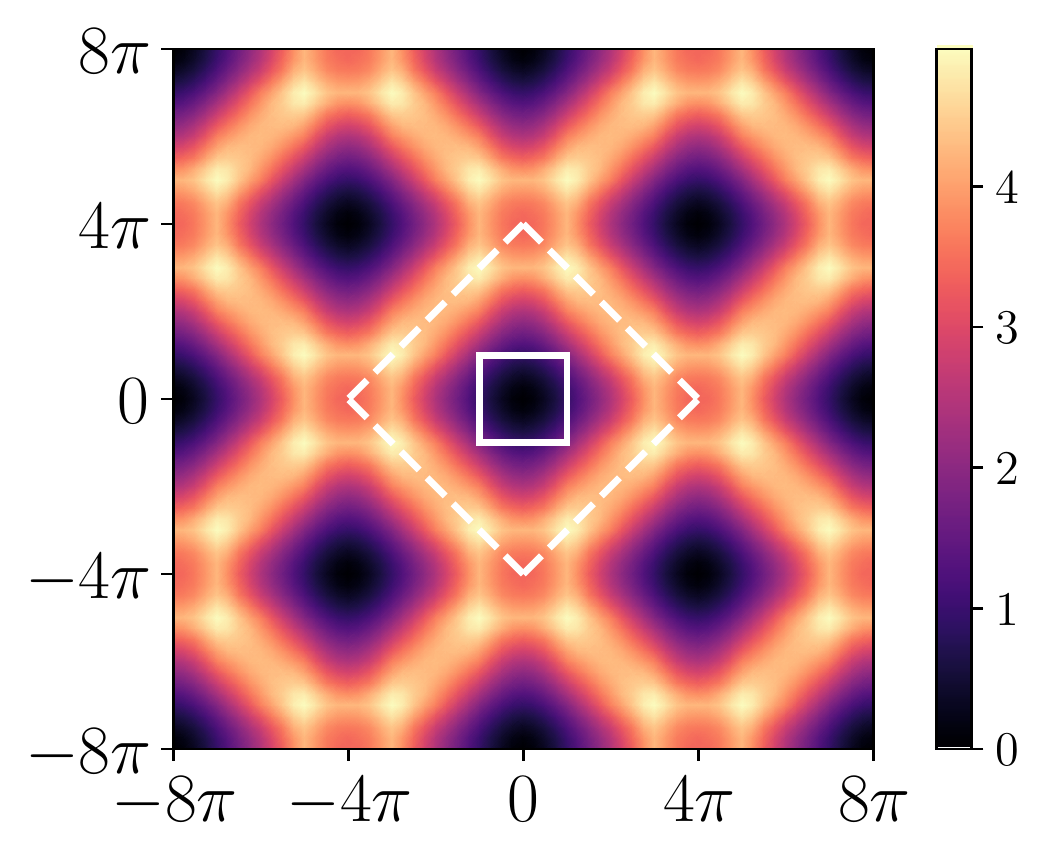}};
        \node at (0.12, 1.9)  {(a) VMC};
        \node at (4.15, 1.9)  {(b) mVMC};
    \end{tikzpicture}
    \caption{The static (equal-time) spin structure factor $S(\mathbf q)$ obtained within (a) VMC and (b) mVMC on the $L=8$ cluster. The solid (dashed) white lines mark the first (extended) Brillouin zones.}
    \label{fig:Sq_8x8}
\end{figure}

{\it Conclusions}. We have employed state-of-the-art numerical quantum many-body methods to provide compelling evidence that the ground state of the $S=1/2$ Heisenberg antiferromagnet on the {\it shuriken} lattice features long-range dimer-order breaking translational invariance, i.\,e., a VBC. Combining the two variational methods, (i) VMC with {\it a priori} given different QSL and VBC Ans\"atze, and (ii) mVMC involving an unconstrained optimization of the projected-BCS wave function, we have revealed a consistent picture of a VBC with pinwheel structure of correlations as inferred from a comprehensive analysis of the dimer-dimer correlation function. This finding is at variance with that obtained within an extended (beyond nearest-neighbor valence bond basis) quantum dimer model framework which argued for a loop-6 VBC~\cite{Ralko-2015}. Given that KCu$_6$AlBiO$_4$(SO$_4$)$_5$Cl~\cite{Fujihala-2020} realizes a gapless spin liquid, and consideration of a generalized model with $J_{\bigtriangleup}\neq J_{\square}$ and further neighbor couplings fails to reproduce the neutron scattering profile as shown in Ref.~\cite{Fujihala-2020}, possibly hints at the role of nonnegligible Dzyaloshinskii-Moriya interactions at play, and the investigation of these interactions would constitute an important future direction of research. Finally, given that (lattice) nematic topological quantum spin liquids have been proposed as competitive Ans\"atze~\cite{Lugan-2019}, a projective symmetry group classification~\cite{wen2002s} of fermionic mean-field Ans\"atze of symmetric and nematic $\mathbb{Z}_{2}$ spin liquids, and a subsequent analysis of the energies and correlation functions of the corresponding Gutzwiller projected spin states would constitute an important direction for future investigations.

{\it Acknowledgments}. We thank Arnaud Ralko, Ludovic Jaubert, Atanu Maity, and Federico Becca for illuminating discussions. Y.\,I. acknowledges financial support by the Science and Engineering Research Board (SERB), Department of Science and Technology (DST), India through the Startup Research Grant No.~SRG/2019/000056, MATRICS Grant No.~MTR/2019/001042, and the Indo-French Centre for the Promotion of Advanced Research (CEFIPRA) Project No. 64T3-1. This research was supported in part by the National Science Foundation under Grant No.~NSF~PHY-1748958, the Abdus Salam International Centre for Theoretical Physics (ICTP) through the Simons Associateship scheme funded by the Simons Foundation, IIT Madras through the Institute of Eminence (IoE) program for establishing the QuCenDiEM group (Project No. SB20210813PHMHRD002720) and FORG group (Project No. SB20210822PHMHRD008268), the International Centre for Theoretical Sciences (ICTS), Bengaluru, India during a visit for participating in the program “Novel phases of quantum matter” (Code: ICTS/topmatter2019/12). Y.\,I. acknowledges the use of the computing resources at HPCE, IIT Madras. F.\,F. acknowledges support from the Alexander von Humboldt Foundation through a postdoctoral Humboldt fellowship. 
N.\,A is funded by the Swiss National Science Foundation, grant number: PP00P2{\_}176877. The mVMC simulations (results section) were supported by the RSF grant (project No.\,21-12-00237). The authors acknowledge the usage of computing resources of the federal collective usage center ``Complex for simulation and data processing for mega-science facilities'' at NRC ``Kurchatov Institute'', \href{http://ckp.nrcki.ru}{http://ckp.nrcki.ru}.
N.\,N. acknowledges funding from the German Research Foundation within the TR 183 (project A04) and the usage of computing resources at the Curta cluster provided by the ZEDAT \cite{Bennett2020}. The work in Würzburg was supported by the Deutsche Forschungsgemeinschaft (DFG, German Research Foundation) through Project-ID 258499086-SFB 1170 and the Würzburg-Dresden Cluster of Excellence on Complexity and Topology in Quantum Matter – ct.qmat Project-ID 390858490-EXC 2147.
T.\,M., P.\,G., and R.\,T. gratefully acknowledge the Gauss Centre for Supercomputing e.\,V. (\href{www.gauss-centre.eu}{www.gauss-centre.eu}) for funding this project by providing computing time on the GCS Supercomputer SuperMUC at Leibniz Supercomputing Centre (\href{www.lrz.de}{www.lrz.de}).


%%%%%%%%%%%%%%%%%%%%%%%%%%%%%%%%%%%%%%

% Supplemental Material

\newcommand{\beginsupplement}{%
        \setcounter{table}{0}
        \renewcommand{\thetable}{S\arabic{table}}%
        \setcounter{figure}{0}
        \renewcommand{\thefigure}{S\arabic{figure}}%
        \setcounter{equation}{0}
        \renewcommand{\theequation}{S\arabic{equation}}%
        \setcounter{page}{1}
     }
 
 \bibliographystyle{apsrev4-1}

\begin{thebibliography}{29}
\bibitem [{not()}]{note_peaks}%
  \BibitemOpen
  \href@noop {} {}\bibinfo {note} {Additional subdominant Bragg peaks, carrying
  (i) 12\% of the spectral weight of the dominant ones are located at $(2\pi -
  q, q)$, $(q, 2\pi - q)$, $(-(2\pi - q), 2q)$ and $(2q, -(2\pi - q))$ and (ii)
  carrying 4\% intensity at $(-(2\pi - q), 2\pi - q)$, $(q, 2 q)$ and $(2 q,
  q)$.}\BibitemShut {Stop}%
\end{thebibliography}
\renewcommand*{\citenumfont}[1]{S#1}
\renewcommand*{\bibnumfmt}[1]{[S#1]}
 
\newcommand\blankpage{%
    \null
    \thispagestyle{empty}%
    \addtocounter{page}{-1}%
    \newpage}

\blankpage
\blankpage

\chead{{\large \bf{---Supplemental Material---}}}

\thispagestyle{fancy}

\beginsupplement

\maketitle

\section{Lattice Convention}\label{sec:latconv}
\begin{figure}[b]
	\centering
	\includegraphics[width=0.6\columnwidth]{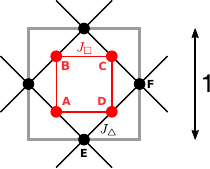}
	\caption{A single unit cell of the \textit{shuriken} lattice. Different colors refer to symmetry-inequivalent sublattice sites ($A,B,C,D$ in red, $E,F$ in black) and symmetry-inequivalent bonds 
	($J_{\square}$ in red, $J_{\bigtriangleup}$ in black). Our study focuses on the highly-frustrated antiferromagnetic Heisenberg Hamiltonian with $J_{\square}=J_{\bigtriangleup}=J$.
	\label{fig:lattice}}
\end{figure}

The {\it shuriken} lattice is defined by a square Bravais lattice with unit vectors 
\begin{equation}
    \mathbf{a}_1 = (1,0) \quad \mathbf{a}_2 = (0,1),
\end{equation}
and a unit cell of six sublattice sites, labelled from $A$ to $F$, whose relative coordinates are
\begin{align}
    \boldsymbol{\delta}_A &= (-1/4,-1/4)& 
    \boldsymbol{\delta}_B &= (-1/4,1/4) \nonumber \\
    \boldsymbol{\delta}_C &= (1/4,1/4) &
    \boldsymbol{\delta}_D &= (1/4,-1/4) \nonumber \\ 
    \boldsymbol{\delta}_E &= (0,-1/2)& 
    \boldsymbol{\delta}_F &= (1/2,0)
\end{align}
The resulting network of sites outlines a {\it shuriken} shape inside the unit cell, made of four corner-sharing triangles surrounding a square, as shown in Fig.~\ref{fig:lattice}. The sublattice sites can be split into two topologically distinct classes, one formed by $A,B,C$ and $D$ sites (vertices of the square), and one made of $E$ and $F$ sites (vertices of the spikes of the {\it shuriken}). The sublattice sites inside each class are transformed into each other by lattice symmetries. We note that in our convention all the sites coordinates are commensurate, and, therefore, an extended Brillouin zone can be defined in reciprocal space, delimited by the square connecting the $(\pm 4\pi,0)$ and $(0,\pm 4\pi)$ momenta.

\section{Methods}

\subsection{Large-$N$ approximation}
\begin{figure}
	\centering
	\includegraphics[width=1.0\columnwidth]{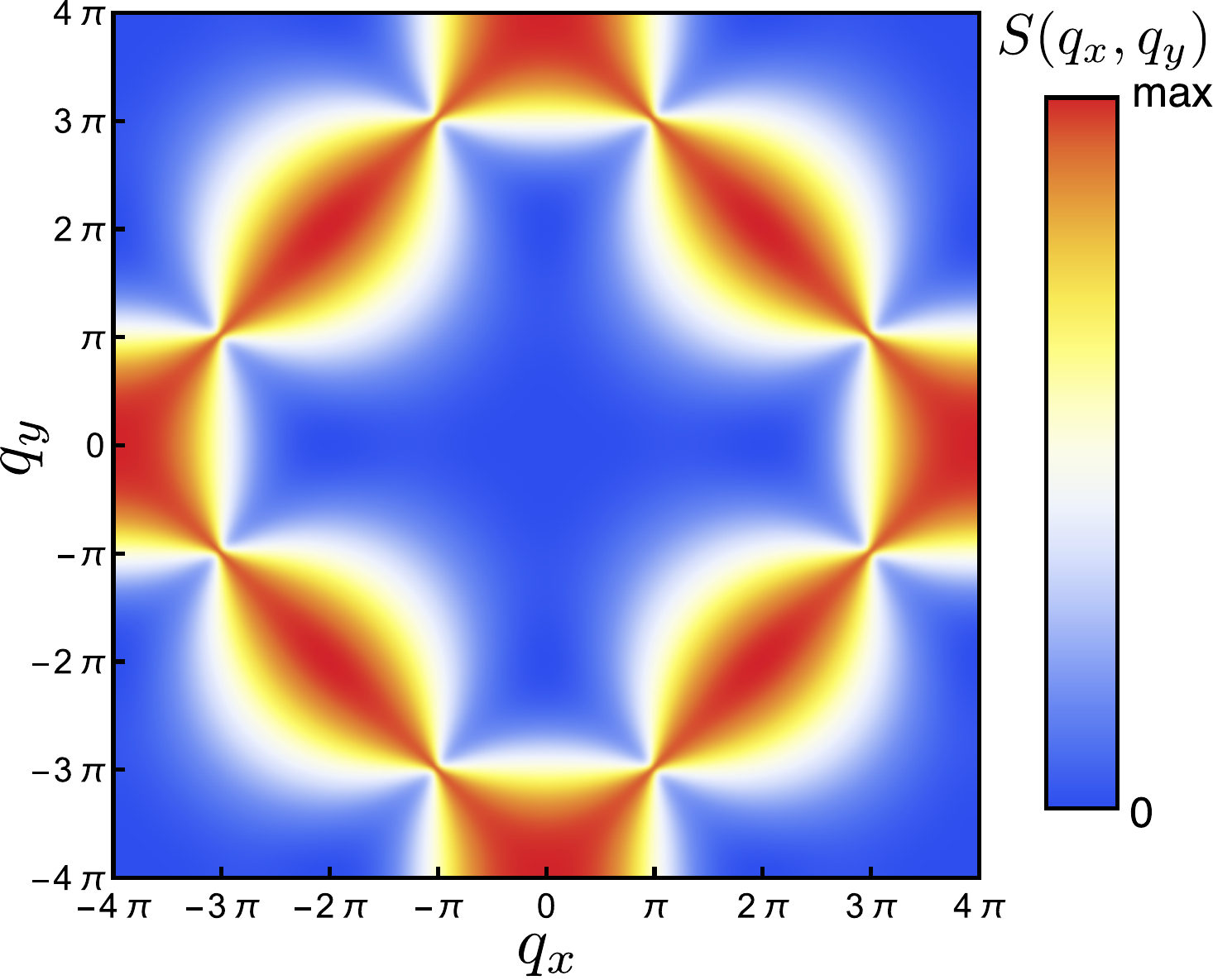}
	\caption{Static structure factor obtained via large-$N$ analysis featuring pinch-points at the boundary of the extended Brillouin zone.} \label{fig:large_n}
\end{figure}

In Fig.~\ref{fig:large_n}, we show the static (equal-time) spin structure factor (at $T=0$) for the Heisenberg model on the \textit{shuriken} lattice as obtained by a standard large-$N$ approximation~\cite{isakov04}. This quantity is defined as 
\begin{equation}
    \label{eq:S_q}
    S(\mathbf q) = \frac{1}{N_{\mbox{\small s}}} \sum\limits_{0 \leqslant i,\,j < N_{\mbox{\small  s}}} \langle \hat{\mathbf{S}}_i \cdot \hat{\mathbf{S}}_j \rangle e^{\imath \mathbf q\cdot (\mathbf r_i - \mathbf r_j)},
\end{equation}
where $N_{\mbox{\small  s}}$ is the number of sites in the lattice, $\mathbf q$ is a momentum inside the extended Brillouin zone, and $\mathbf r_i$ denotes the site positions (accounting for sublattice displacements), following the convention outlined in Section~\ref{sec:latconv}.
Within the large-$N$ approach, the number of components of a classical spin vector is artificially increased from three to infinity and the resulting model is treated in the saddle-point approximation. The spin structure factor shows eight pinch point singularities inside the extended Brillouin zone, which are characteristic for classical spin models with corner-sharing triangles. In these models, the ground states are determined by the condition that spins sum up to zero in each triangle. Since these conditions can be fulfilled by an extensive number of spin configurations the resulting state is highly degenerate and referred to as a classical spin liquid. The fact that spin correlations in this state fall off with distance according to a power law, is tightly connected with the existence of pinch point singularities.

\subsection{Variational Monte Carlo (VMC)} \label{sec:VMC}

To unveil the ground state properties of the Heisenberg model on the \textit{shuriken} lattice we resort to a variational Monte Carlo (VMC) approach~\cite{becca_quantum_2017}. 
Our variational \textit{Ans\"atze} consist of uncorrelated fermionic wave functions (i.e., Slater determinants), which are constrained to the spin space by Gutzwiller projection. The construction of the variational states relies on the Abrikosov fermion representation of spin degrees of freedom, in which $S=1/2$ spin operators are rewritten as suitable combinations of fermionic bilinears, i.e.,
\begin{equation}\label{eq:Sabrikosov}
\hat{\mathbf{S}}_i=\frac{1}{2}\sum_{\alpha,\beta=\downarrow,\uparrow} \hat{c}_{i,\alpha}^{\dagger} \boldsymbol{\sigma}_{\alpha,\beta} \hat{c}_{i,\beta}.
\end{equation}
Here, $\hat{c}_{i,\alpha}$ /  $\hat{c}_{i,\alpha}^\dagger$ are fermionic operators which annihilate/create a fermion with spin $\alpha$ ($=\uparrow $ or $\downarrow$) at site $i$, and  $\boldsymbol{\sigma}=(\sigma^x,\sigma^y,\sigma^z)$ is a vector of Pauli matrices. 
The Abrikosov representation preserves the algebra of the spin operators, but artificially enlarges the Hilbert space of the system, introducing spurious fermionic states with empty and doubly occupied sites. Yet, this fermionic framework can be exploited for the definition of variational wave functions for the Heisenberg model, provided that a projection onto the spin Hilbert space is performed. In other words, we can define variational wave functions for spins as
\begin{equation}
|\Psi_0\rangle=\hat{\mathcal{P}}^{\infty}_{\mbox{\footnotesize G}} |\Phi_0\rangle,
\end{equation}
where $|\Phi_0\rangle$ is a certain uncorrelated fermionic state (see below) and $\hat{\mathcal{P}}^{\infty}_{\mbox{\footnotesize G}}$ is the Gutzwiller projection operator, 
\begin{equation}
\hat{\mathcal{P}}^{\infty}_{\mbox{\footnotesize G}}=\prod_i (\hat{c}_{i,\uparrow}^{\dagger}\hat{c}_{i,\uparrow}-\hat{c}_{i,\downarrow}^{\dagger}\hat{c}_{i,\downarrow})^2,
\end{equation}
which enforces single fermionic occupation of lattice sites. In this work, we choose $|\Phi_0\rangle$ to be the ground state of a quadratic fermionic Hamiltonian, dubbed $\hat{\mathcal{H}}_{\rm 0}$. For non-magnetic phases we take a tight-binding Hamiltonian
\begin{equation}\label{eqn:mf-nonmag}
\hat{\mathcal{H}}_{\rm 0}=\sum_{(i,j),\alpha}\chi_{ij}\hat{c}_{i,\alpha}^{\dagger}\hat{c}_{j,\alpha},
\end{equation}
with nearest-neighbor hopping terms, $t_{i,j}$, which play the role of variational parameters. An appropriate parametrization of $\hat{\mathcal{H}}_{\rm 0}$, based on a projective symmetry group (PSG) analysis~\cite{wen2002,bieri2016}, allows us to define different U(1) spin-liquid \textit{Ans\"atze} on the \textit{shuriken} lattice, both fully symmetric and chiral, which are summarized in Fig.~\ref{fig:ansatze}. The various spin-liquid states are distinguished by different gauge-invariant fluxes threading the elementary plaquettes of the lattice. These fluxes are determined by the
signs/phases of the hoppings, which are fixed by the projective symmetry group of the \textit{Ansatz}~\cite{wen2002,bieri2016}. On the other hand, the amplitudes of the hoppings are free variational parameters, which are optimized to minimize the expectation value of the energy. Specifically, the fermionic Hamiltonian defining the spin liquid states depends on the ratio of two hopping amplitudes, one for the bonds forming the squares (i.e., the bonds connecting sites of sublattices $A,B,C,D$), named $\chi_{\square}$, and one for the bonds forming the triangular spikes of the \textit{shurikens} (i.e., the bonds connecting sites of sublattices $A,B,C,D$ to sites of sublattices $E,F$), named $\chi_{\bigtriangleup}$.

As discussed in the main text, some of the spin-liquid states of Fig.~\ref{fig:ansatze} are unstable to the formation of valence-bond order. This means that a variational energy gain is achieved by letting the hopping amplitudes take different values on different bonds~\cite{iqbal2012,ferrari2017}, according to certain dimer patterns~\cite{Rousochatzakis-2013}, as outlined in the main text. In particular, a pinwheel valence-bond crystal wave function is found to provide the best variational energy for the Heisenberg model on the \textit{shuriken} lattice. The pinwheel variational state explicitly breaks lattice translations, requiring a $2\times 2$ enlargement of the unit cell, and has a lower point group symmetry ($D_2$).

Additional VMC calculations are performed to rule out the possible presence of long-range magnetic order in the ground state of the system. For this purpose, we define magnetically ordered variational wave functions by introducing a fictitious Zeeman field in the fermionic auxiliary Hamiltonian, i.e.,
\begin{equation}\label{eqn:mf-mag}
\hat{\mathcal{H}}_{\rm mag}=\sum_{(i,j),\alpha}\chi_{ij}\hat{c}_{i,\alpha}^{\dagger}\hat{c}_{j,\alpha} +
h_{\rm AF} \sum_i \mathbf{m}_i \cdot\hat{\mathbf{S}}_i,
\end{equation}
Here, the variational parameter $h_{\rm AF}$ determines the magnitute of the Zeeman field, while the unit vector $\mathbf{m}_i$ modulate its local orientation. We consider a coplanar field in the $xy$-plane, taking ${\mathbf{m}_i =[\cos(\mathbf{q}\cdot \mathbf{R}_i+\theta_i),\sin(\mathbf{q}\cdot \mathbf{R}_i+\theta_i),0]}$, where $\mathbf{R}_i$ is the coordinate of the unit cell of site $i$, $\mathbf{q}$ is the ordering vector and $\theta_i$ are sublattice-dependent angles. By appropriately choosing $\mathbf{q}$ and $\theta_i$, we can impose different magnetic orders in the wave function. As discussed in the main text we consider two possible ordering patterns, with neighboring spins oriented with a relative angle of $120^\circ$. For what concerns the hopping terms of $\hat{\mathcal{H}}_{\rm mag}$, we take a uniform \textit{Ansatz}, with $\chi_{\square}$ ($\chi_{\bigtriangleup}$) being the hopping on the bonds forming the squares (triangular spikes) of the \textit{shurikens}. The variational wave function for these magnetic phases is then obtained by projecting the ground state of Eq.~\eqref{eqn:mf-mag}, dubbed $|\Phi_{\rm mag}\rangle$, 
\begin{equation}
|\Psi_{\rm mag}\rangle=\hat{\mathcal{J}}\hat{\mathcal{P}}^{\infty}_{\mbox{\footnotesize G}} |\Phi_{\rm mag}\rangle.
\end{equation}
In addition to the Gutzwiller-projection, we introduce a long-range spin-spin Jastrow factor, 
\begin{equation}
\hat{\mathcal{J}}=\exp\left(\sum_{i,j} v_{i,j} \hat{S}^z_i \hat{S}^z_j \right)
\end{equation}
which induces additional transverse quantum fluctuations on top of the magnetic order in the $xy$-plane. The pseudopotentials $v_{i,j}$ are taken to be translationally invariant and fully optimized to minimize the variational energy, together with the parameters of $\hat{\mathcal{H}}_{\rm mag}$, namely $\chi_{\bigtriangleup}$ and $h_{\rm AF}$ (we set $\chi_{\square}=1$ to fix the overall energy scale of the auxiliary Hamiltonian, which is irrelevant). After optimization, a finite (zero) value of $h_{\rm AF}$ in the thermodynamic limit indicates the presence (absence) of long-range magnetic order.

Our variational calculations are performed on $6\times L \times L$ clusters with periodic boundary conditions which respect the full symmetry of the {\it shuriken} lattice. For the numerical optimization of the variational parameters we apply the stochastic reconfiguration method~\cite{sorella_green_1998}.

Once we have obtained the optimal variational wave function $|\Psi_0\rangle$, which corresponds to the pinwheel VBC state, we can iteratively apply a number of Lanczos steps to systematically improve the variational energy~\cite{laflorencie2004,sorella2001,becca2015}. This procedure yields the variational state
\begin{equation}
|\Psi_p\rangle = \left ( \sum_{n=0}^p \alpha_n \hat{\mathcal{H}}^n \right )|\Psi_0\rangle,
\end{equation}
after $p$ iterations, where the coefficients $\alpha_n$ are variational parameters that minimize the expectation value of the energy
\begin{equation}
E_p = \langle \Psi_p|\hat{\mathcal{H}}|\Psi_p\rangle.
\end{equation}
We perform $p=0,1,2$ Lanczos steps and compute an estimate of the energy in the $p\rightarrow \infty$ limit by extrapolating $E_p$ as a function of the variance~\cite{sorella2001,becca2015}
\begin{equation}\label{eqn:variance}
\sigma^2_p = \langle \Psi_p|\hat{\mathcal{H}}^2|\Psi_p\rangle - \langle \Psi_p|\hat{\mathcal{H}}|\Psi_p\rangle^2.
\end{equation}

\subsection{Many-variable wave function (mVMC)}
\label{sec:mVMC}
The many-variable variational Monte Carlo (mVMC) method can be successfully used in studies of strongly correlated spin and electronic systems~\cite{PhysRevB.90.115137,casula2004correlated}. In particular, the method can be applied to distinguish between quantum spin liquid and valence bond solid phases, such as in the case of the $j_1$-$j_2$ Heisenberg model on the square lattice~\cite{doi:10.7566/JPSJ.84.024720,nomura2020dirac}. In this work, we employ the mVMC implementation from Ref.~\cite{misawa2019mvmc,doi:10.1143/JPSJ.77.114701}. To apply mVMC, fermionic operators are first mapped to pseudofermionic bilinears as given in Eq.\,\eqref{eq:Sabrikosov}.

\begin{figure}
  \includegraphics[width = \linewidth]{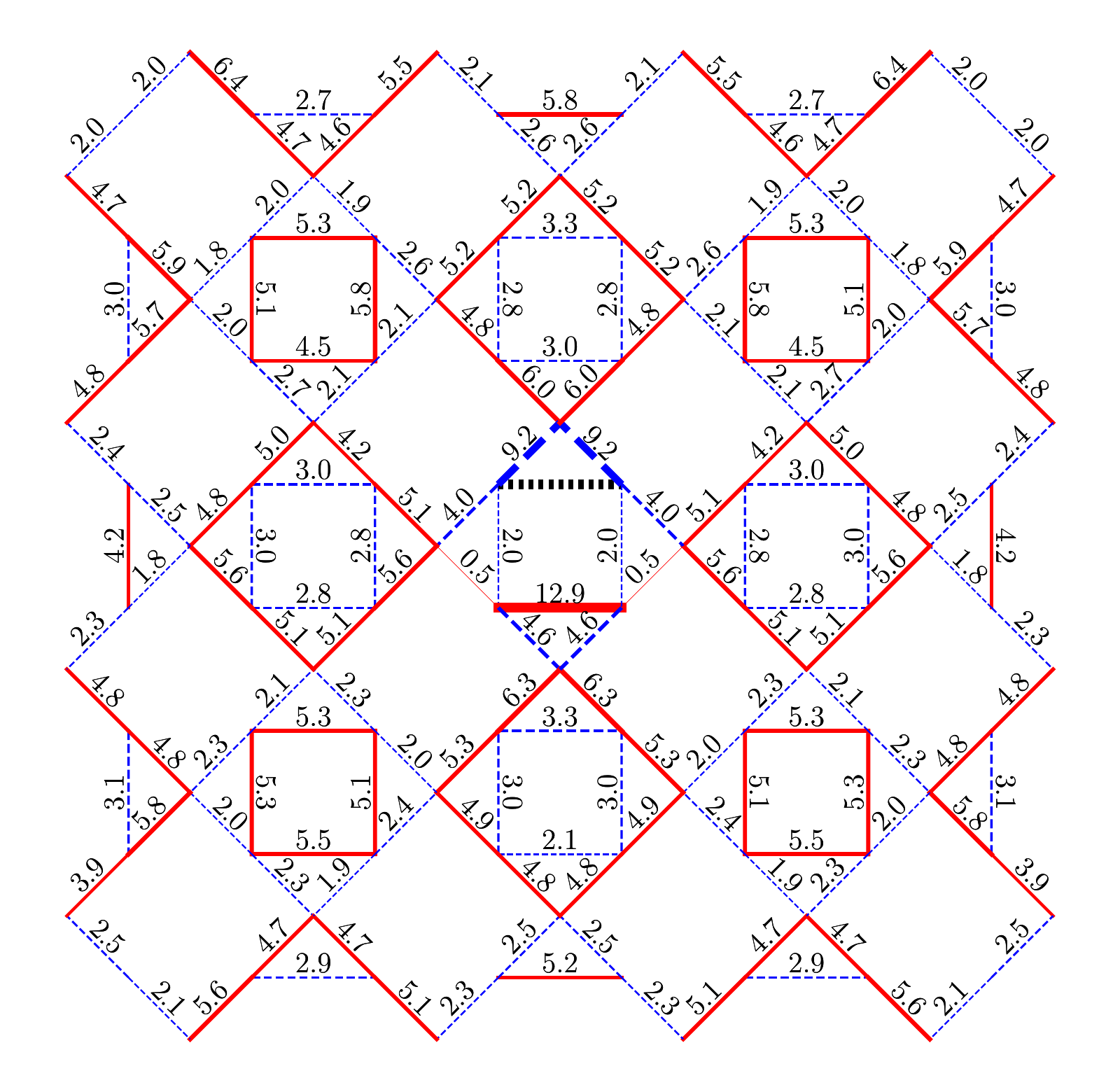}\\\includegraphics[width = \linewidth]{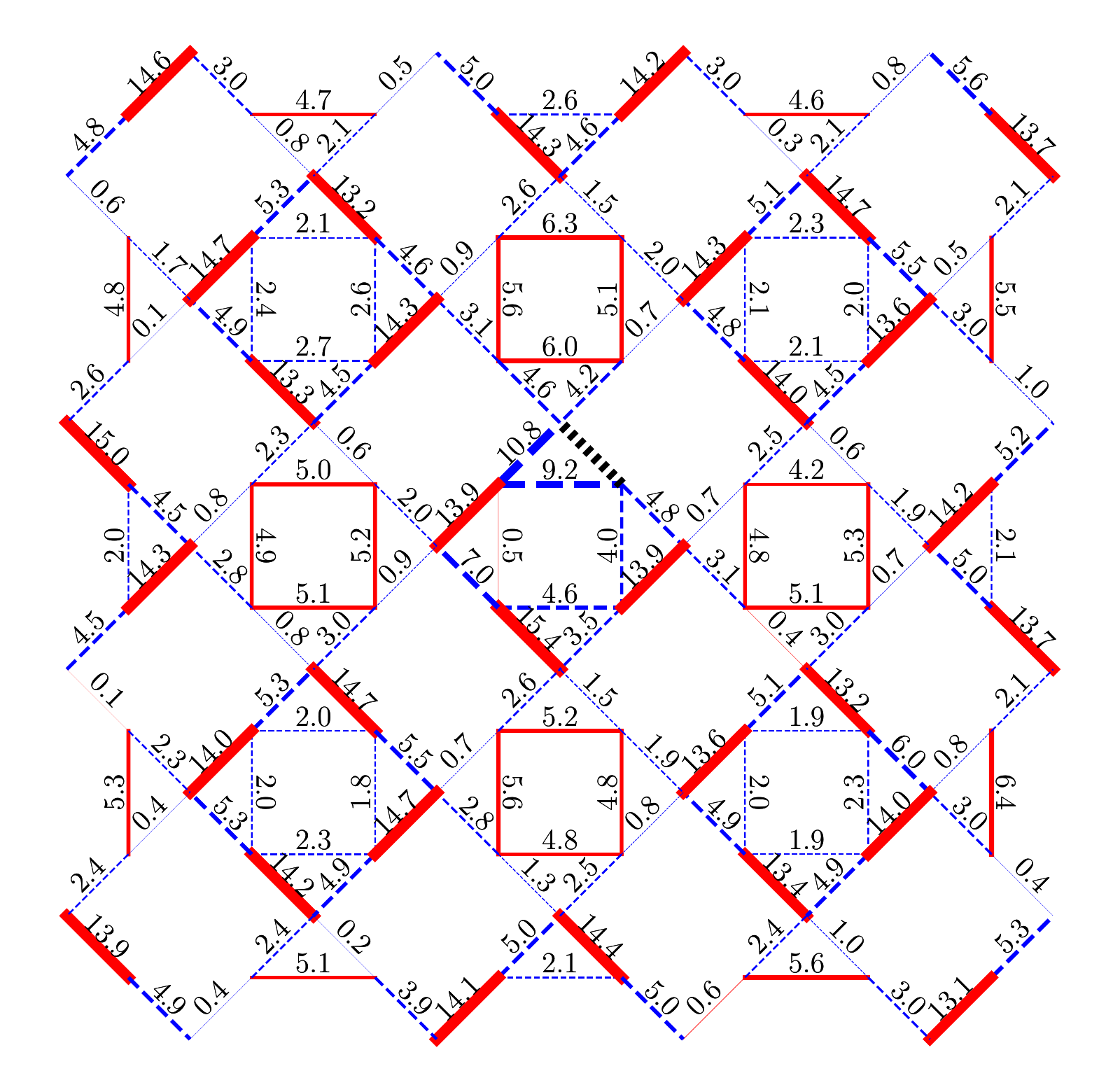}
  \caption{Dimer-dimer correlations between the base bond, marked by a dotted black line, and other bonds measured within mVMC on the $L=8$ cluster. 
    (for typographical reasons $\chi_{b,b'}^{D}$ is multiplied by $10^2$ and rounded to 2 digits). 
    Solid red (dashed blue) lines represent positive (negative) correlation values. {\it Top panel}: base bond is between sublattices B and C. {\it Bottom panel}: base bond is between sublattices C and E.}
  \label{fig:mvmc_dimer_both}
\end{figure}

Inspired by the Anderson resonating valence-bond wave function, the mVMC ansatz has the form
\begin{equation}
    |\phi_{\mbox{\footnotesize pair}} \rangle = \hat{\mathcal{P}}^{\infty}_{\mbox{\footnotesize G}} \exp \left(\sum\limits_{i, j} f_{i,j} \hat c^{\dagger}_{i, \uparrow} \hat c^{\dagger}_{j \downarrow} \right) |0 \rangle,
\end{equation}
where single occupation is ensured by the $\hat{\mathcal{P}}^{\infty}_{\mbox{\footnotesize G}}$ Gutzwiller projector such that fermionic Hilbert space is mapped to the original Hilbert space of spin operators. The wave-function value $\langle \boldsymbol{\sigma} |\phi_{\mbox{\footnotesize pair}} \rangle $ of a specific spin configuration $|\boldsymbol{\sigma}\rangle$ is evaluated using the Slater determinant of the matrix with elements $f_{i,j}$. Here, $\boldsymbol{\sigma}$ represents a string of $\pm 1$, which, for each lattice site,  stands for the respective spin eigenstate in the $S^z$ basis. The parameters $f_{i,j}$ are optimised using the stochastic reconfiguration optimization technique~\cite{sorella_green_1998}, which can be seen as a way of performing imaginary-time evolution in the variational parameters manifold~\cite{becca_quantum_2017,carleo2017solving}.

To obtain accurate wave functions, we employ quantum-number projections. The point-group symmetry $\hat G$ is projected by applying its generators until the symmetry orbit is exhausted
\begin{equation}
 |\Psi_\xi \rangle = \hat P |\Psi \rangle =  \sum\limits_n \xi^n \hat G^n |\Psi \rangle,
    \label{eq:q_projection}
\end{equation}
where $\xi$ is the desired projection quantum number and $|\Psi_\xi \rangle$ the resulting symmetrized state. The projection onto the total spin $S$ is performed by superposing the $SU(2)$--rotated wave functions~\cite{doi:10.1143/JPSJ.77.114701}. In this work, for systems with more than 36 sites, we partially impose translational symmetry directly on the variational parameters $f_{i,\,j}$. Namely, we introduce translational symmetry modulo $2 \times 2$ unit cells sublattice structure and project the $2 \times 2$ translations and the point-group symmetries using Eq.\,\eqref{eq:q_projection}. The resulting procedure amounts into $2 \times 2 \times 6^2 \times L^2$ variational parameters with $L$ being the number of unit cells in each lattice direction. Such partial translational symmetry imposition is a reasonable compromise between the ability to express complicated wave function and and the required time to learn the wave function.

In Fig.\,\ref{fig:mvmc_dimer_both} we show dimer-dimer correlations measured within mVMC on the $8 \times 8$ lattice.

\section{PFFRG}
A powerful and unbiased method in the investigation of the interacting many body problem is found in the functional renormalization group (FRG) which reformulates such problem in terms of an infinite hierarchy of first order differential equations for the $n$-particle vertex functions~\cite{Kopietz2010}. While the fermionic FRG formalism is well-defined for models of Hubbard-type~\cite{Halboth2000,Classen2020}, employing spin operators directly in an FRG formalism is fundamentally challenging and requires significant alteration of the usual method~\cite{Tarasevych2021}. However, using a decomposition of spin operators into pseudofermions as in Eq.~\ref{eq:Sabrikosov}, one can map this problem onto the fermionic Hilbert space instead. The problem of unphysical states introduced by this parton-decomposition of spins is circumvented by a restriction to zero temperatures, where these states are expected to gap out due to their vanishing spin $S=0$. 
In this \textit{pseudofermion (PF-)} FRG, the bare pseudofermion propagator is initially modified by the introduction of a sharp Matsubara frequency cutoff $G^0_i(\omega)\rightarrow \theta(|\omega| - \Lambda) G^0_i(\omega)$. This cutoff defines the renormalization scale $\Lambda$ which enables turning from an effective high-energy model at $\Lambda \rightarrow \infty$, in which the pseudofermion vertices are trivially known, to the desired cutoff-free limit at $\Lambda = 0$. To resolve the $\Lambda$-dependence of vertices the emerging infinite set of coupled integro-differential equations is truncated at the three-particle level. The obtained finite number of the so-called flow equations for the self-energy and two-particle vertex may then be integrated numerically. Most importantly, the PFFRG employs the improved Katanin truncation scheme, which includes crucial contributions from the three-particle vertex \cite{Katanin2004}. Using this scheme, the FRG becomes asymptotically exact in the limit of large spin magnitudes $S$ and large $N$, where $SU(N)$ refers to the spin's symmetry group \cite{Baez2017,Buessen2018}. Since the limits $S \rightarrow \infty$ and $N \rightarrow \infty$ capture the physical properties of magnetically ordered and disordered states, respectively, the PFFRG has achieved great success in mapping out the magnetic phase diagrams of a variety of two- and three-dimensional systems. 
From the two-particle vertex, any correlator bilinear in the fermionic creation and annihilation operators can be recovered, most importantly, the static magnetic susceptibility defined as

\begin{equation}
    \chi^\Lambda_{ij} = \int_0^\infty d\tau \langle S^z_i(\tau) S^z_j(0) \rangle.
\end{equation}

While at larger cutoff scales the susceptibility is suppressed, at low $\Lambda \lesssim J$ instabilities can be observed at the onset of spontaneous magnetic ordering. Since correlations in the two-particle vertices can only be treated numerically up to a maximal correlation distance, the onset of magnetic order is featured by no divergence, but rather a finite peak of the Fourier-transformed susceptibility $\chi_k^\Lambda$
at a critical energy scale $\Lambda_c \sim T_c$ \cite{Reuther2010,Baez2017}. The point $k$ in momentum space at which this divergence occurs, further specifies the type of magnetic order. Accordingly, magnetically disordered phases display no such feature but rather a continuously growing susceptibility down to $\Lambda = 0$ at all $k$.

\begin{figure*}[t!]
\includegraphics[width=1.0\linewidth]{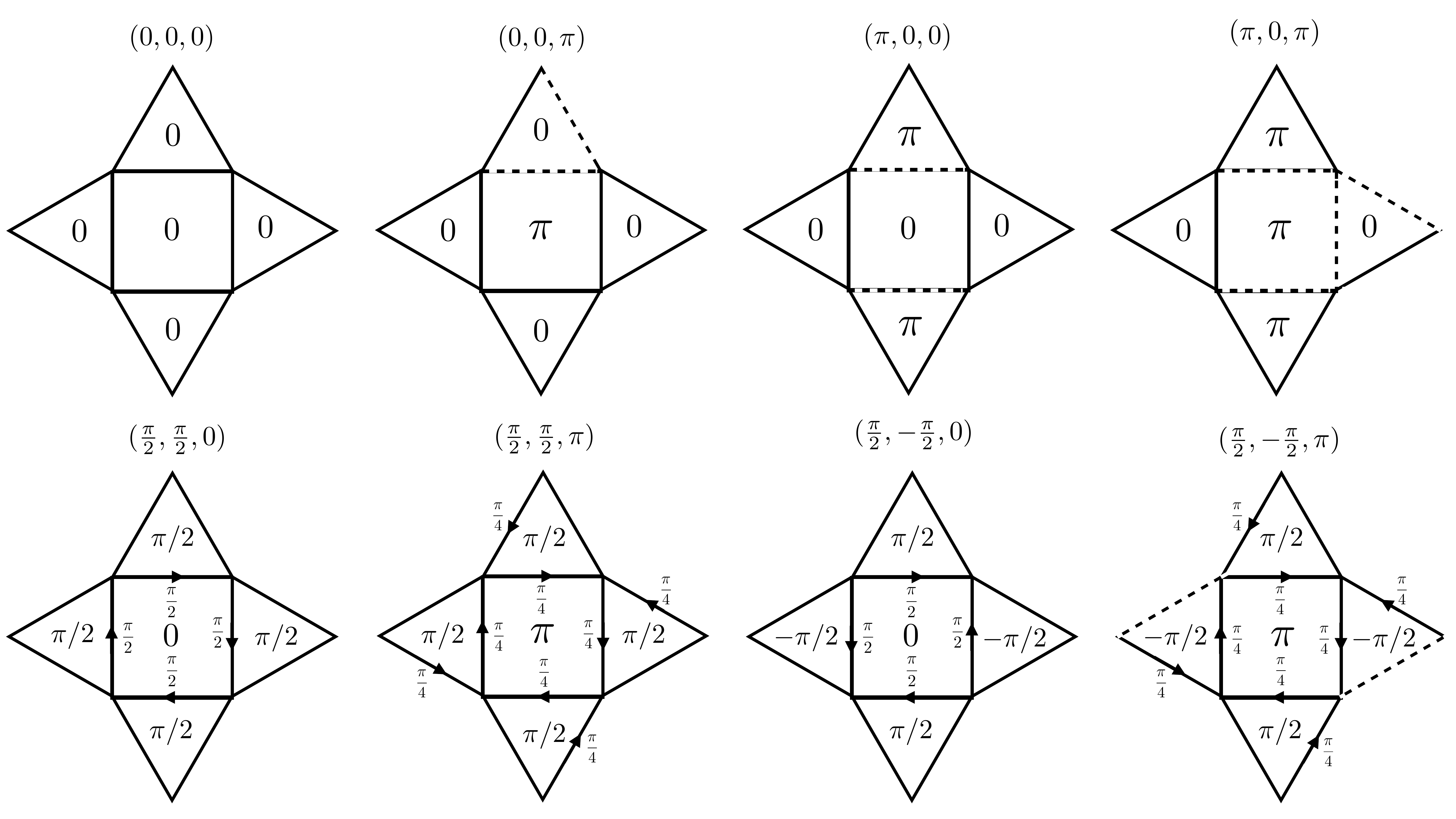}
\caption{The fermionic mean-field Ans\"atze of the $U(1)$ fully symmetric (top row) and chiral (bottom row) quantum spin liquids. The Ans\"atze are completely characterized by gauge invariant fluxes ($\Phi$) through (i) ``up/down'' triangles, (ii) ``right/left'' triangles, (iii) squares, and henceforth, we adopt the notation $(\Phi_{\bigtriangleup},\Phi_{\triangleright},\Phi_{\square})$ to label the Ans\"atze. The solid (dashed) lines represent real positive (negative) hopping, while for the chiral Ans\"atze the nontrivial phases of the bonds are mentioned next to the arrows marking the bond orientation. The ($\frac{\pi}{2}$,$\frac{\pi}{2}$,0) and ($\frac{\pi}{2}$,$\frac{\pi}{2}$,$\pi$) CSLs respect $C_{4}$ rotations ($R$) but break all mirror symmetries ($\sigma$) while preserving the product $\sigma\mathcal{T}$, i.e., they are of the Kalmeyer-Laughlin type. On the other hand, the ($\frac{\pi}{2}$,$-\frac{\pi}{2}$,0) and ($\frac{\pi}{2}$,$-\frac{\pi}{2}$,$\pi$) states breaks both $C_{4}$ rotations and all mirror symmetries while preserving the products $R T$ and $\sigma T$, i.\,e., they are of the staggered flux type.\label{fig:ansatze}}
\end{figure*}

\begin{figure}
  \includegraphics[width = \linewidth]{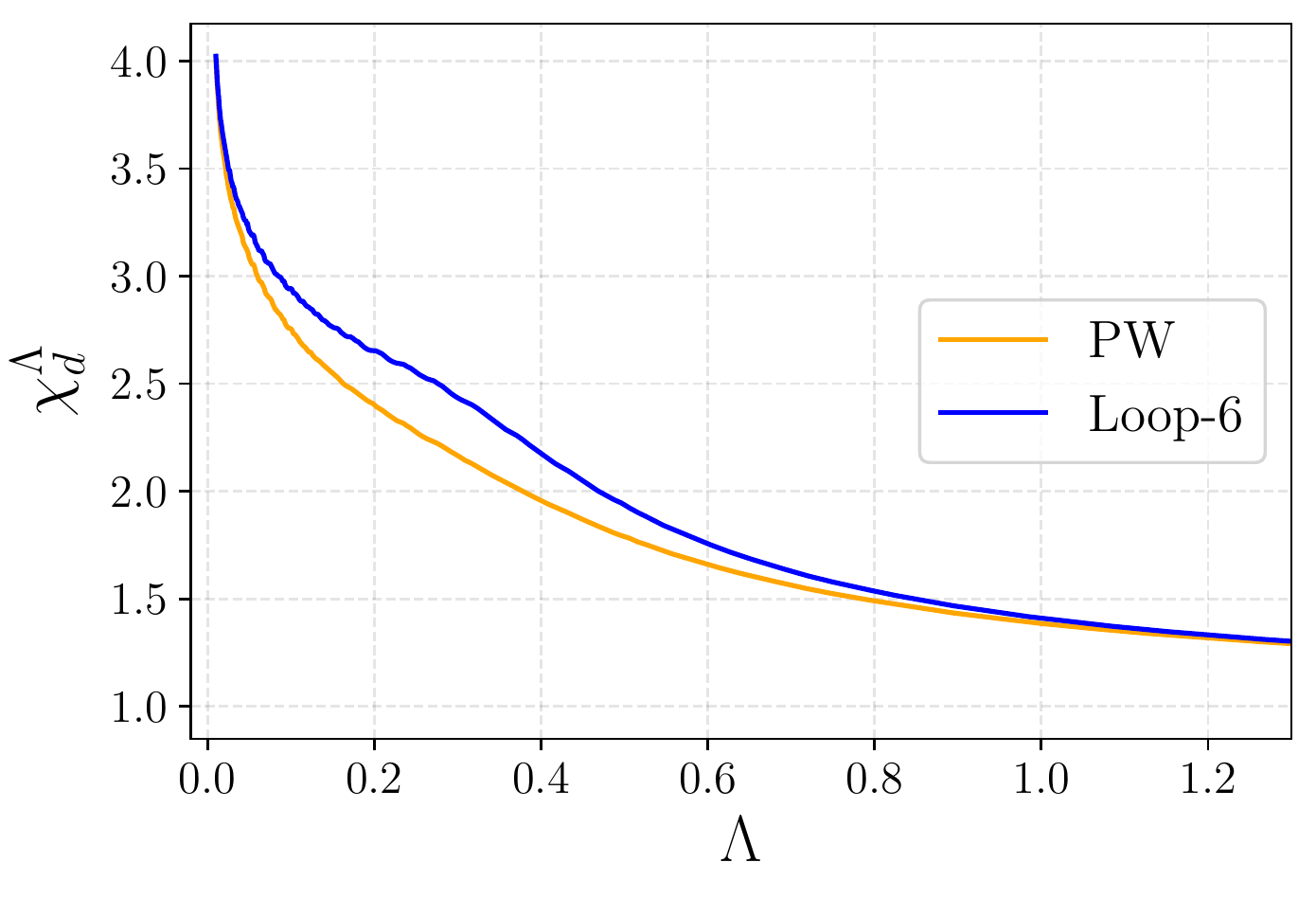}
  \caption{Flow of the dimerization response as defined in Eq.~\ref{eq:pffrg_Dimerization} obtained via PFFRG for the pinwheel (PW) and Loop-6 VBC order, respectively. The results are averaged over pairs of symmetry-inequivalent dimer bonds for easier comparison. A large value of the response at the end of the flow $\chi^{\Lambda = 0}_d \gg 1$ indicates a system's tendency to favor dimerization.}
  \label{fig:pffrg_dimer}
\end{figure}

While dimer- and plaquette correlators themselves are quartic in the spin operators and thus inaccessible to the PFFRG at the current level of truncation, a system's affinity for dimerization can still be investigated for initially specified dimerization patterns. 
To this end, the couplings $J_{ij}$ are initially slightly strengthened by $\delta$ along the dimerized bonds within this pattern, while being weakened on the corresponding non-dimerized bonds~\cite{Iqbal2016}. One may then determine the system's response to this particular dimer pattern as
\begin{equation}
    \chi^\Lambda_d = \frac{J}{\delta} \left| \frac{\chi^\Lambda_+ - \chi^\Lambda_-}{\chi^\Lambda_+ + \chi^\Lambda_-} \right|\text{,} \label{eq:pffrg_Dimerization}
\end{equation}
where $\chi^\Lambda_{+\ (-)}$ refers to the real-space correlator between two spins along such a strengthened (weakened) bond. Dimer states that are accepted by the system typically develop large growth during the flow $\Lambda \rightarrow 0$, while patterns which are rejected remain comparable, or even decrease from their initial value of $\chi_d^{\Lambda \rightarrow\infty} = 1$.
In the present case, we investigate the Pinwheel VBC and the Loop-6 VBC by strengthening the colored bonds in figure 2 of the main text and simultaneously weakening the non-colored bonds by $\delta = 0.01 J$.
Fig.~\ref{fig:pffrg_dimer} shows the RG flows of the dimer response for the L6 and the pinwheel VBC patterns respectively. Since the unit cell allows for the individual evaluation of Eq.~\ref{eq:pffrg_Dimerization} for several nearest neighbor bonds that are distinct by spatial symmetry, we compare the averaged dimer response. In both cases, the average response increases approximately fourfold from its initial value of 1 which indicates a moderate tendency towards dimerization. Although the pinwheel order maintains a slightly stronger response throughout most of the flow, the Loop-6 response remains relatively comparable so that no definite statement regarding the preferred dimerization pattern can be made within this analysis.  

\section{PMFRG}

Instead of implying complex fermions, the FRG formalism can also be applied after representing spin operators through Majorana fermions defined by the anticommutation relation  $ \{ \eta^\alpha_i,\eta^\beta_j \} = \delta_{ij} \delta^{\alpha \beta}$. 
In the $SO(3)$ Majorana representation~\cite{Tsvelik1992}, spin operators are expressed by three different flavors ($x,y,z$) of Majorana fermions
\begin{equation}
  S^x_i = -i \eta^y_i \eta^z_i \text{,} \qquad  S^y_i = -i \eta^z_i \eta^x_i \text{,} \qquad  S^z_i = -i \eta^x_i \eta^y_i \text{.} \label{eq:MajoranaRep}
\end{equation}
Crucially, the Hilbert space in this representation does not contain any unphysical states in strong contrast to the more common complex fermionic representation. This allows for the Pseudo-Majorana (PM-) FRG formalism to be applied at finite temperatures~\cite{Niggemann2021}.

Despite the absence of unphysical states, the $SO(3)$ Majorana Hilbert space is split into many degenerate copies that are only distinguishable by their fermionic parity. Although states of different parity lead to the same physical observables, initially, it was shown that this degeneracy leads to a Curie-type $1/T$ behaviour of certain four-point parity correlators. Not unphysical by itself, the corresponding $1/T$ dependence of certain vertices at very low temperatures shows to affect other observables at the current level of truncation.

\begin{figure}
  \includegraphics[width = \linewidth]{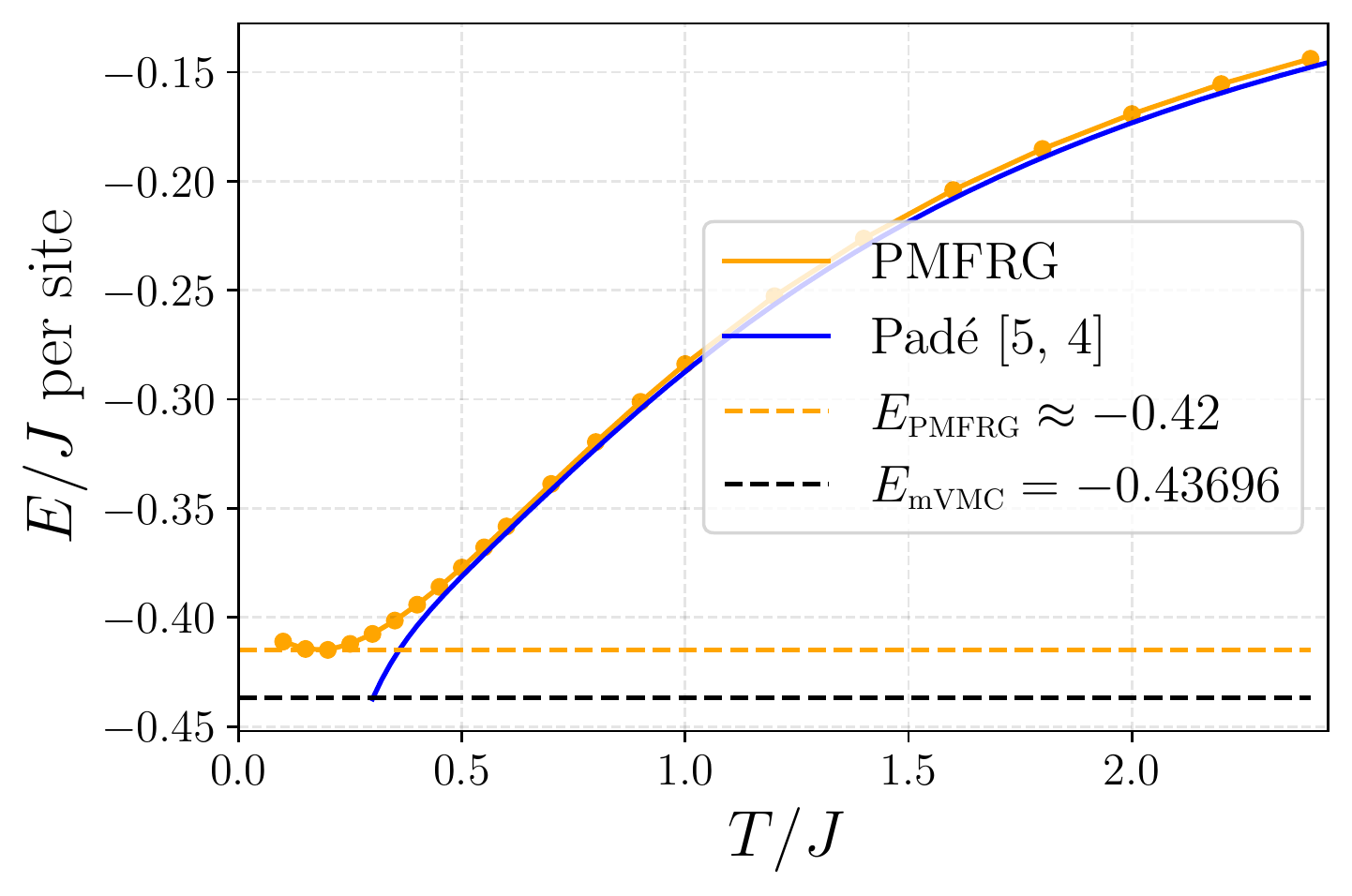}
  \caption{Energy per site as obtained via PMFRG. At higher temperatures, the results can be compared to the Padé $[5,4]$-approximant of the high temperature series expansion to tenth order \cite{Lohmann2014}. $E_\text{PMFRG}$ denotes the lowest energy reached at finite temperatures, whereas $E_\text{mVMC}$ is the many-variable Monte-Carlo result at $T=0$.}
  \label{fig:pmfrg}
\end{figure}

Fig.~\ref{fig:pmfrg} shows the results of a recently developed two-loop PMFRG approach \cite{TwoLoopPMFRG} which extends the method's range of accuracy down to a temperature of $T \sim 0.2  J$. At lower temperatures, strong contributions of divergent vertices lead to an unphysical increase of the energy. We note that in contrast to other approaches taken in this work, the FRG is not motivated from a variational argument and as such the PMFRG energy may in principle be lower than the physical ground state energy, although no such observation was made yet.

\section{\lowercase{i}PEPS}
Infinite Projected Entangled Pair States (iPEPS) are two-dimensional tensor network ansatz, defined directly in the thermodynamic limit~\cite{VerstraetePEPS,JordaniPEPS}. Such techniques do not suffer from negative sign problem and therefore, are very suitable for studying frustrated quantum systems~\cite{Xiangkaf,Picot2015}, and even real materials recently~\cite{KshetrimayumCaCro2020,CorbozSSland2019PRB}.

For this work, we coarse-grain three spins in the original shuriken lattice to a Honeycomb lattice with a two-site unit cell, thus effectively giving us a six-site unit cell. We then employ the simple update~\cite{XiangSU} to optimize the ground state of the model. Similar coarse-graining schemes have been adopted for the Kagome lattice in the past~\cite{Picotkagome2016PRB,KshetrimayumkagomeXXZ2016}. Once the tensors are optimized, the ground state energy is obtained by doing a full contraction of the infinite 2D tensor network using the Corner Transfer Matrix Renormalization Group (CTMRG) algorithm~\cite{ctmnishino1996,ctmnishino1997,ctmroman2009,ctmroman2012}. The ground state energy per site for different bond dimensions of the iPEPS ($D$) and the environment ($\chi$) are shown in the Table \ref{ipepsenergy}. For each bond dimension $D$, the energies are well converged with the environment bond dimension $\chi$.

\begin{table}
\centering
\begin{tabular}{ |p{3cm}||p{3cm}|p{2cm}|  }
 \hline
 \multicolumn{3}{|c|}{Energy per site $E_0$} \\
 \hline
 iPEPS Bond dimension $D$ & Environment bond dimension $\chi$ & $E_0$\\
 \hline
 2   & 4    &-0.408459\\
 -&   10  & -0.408459\\
 3 &9 & -0.419478\\
 -    &15 & -0.419478\\
 4&   16  & -0.425751\\
 -& 30  & -0.425751\\
 5& 10  & -0.428169\\
 -& 25  & -0.42818\\
 6& 20  & -0.430471\\
 -& 36  & -0.430475\\
 7& 49  & -0.431845\\
 -& 80  & -0.431845\\
 8& 64  & -0.432976\\
 -& 80  & -0.432976\\
 9& 81  & -0.433789\\
 -& 100  & -0.433791\\
 \hline
\end{tabular}
\caption{Ground state energy per site $E_0$ for different bond dimensions $D$ of the iPEPS and the environment $\chi$.}
\label{ipepsenergy}
\end{table}

\end{document}